%% file: chauvin-hd196885.tex
\DeclareUnicodeCharacter{0301}{\'{e}} 
\documentclass[longauth]{aa}
\usepackage{graphicx}
\usepackage{txfonts}
\usepackage[citecolor=blue, urlcolor=blue, linkcolor=blue, colorlinks=true]{hyperref}
\usepackage{cleveref}
\usepackage{mathtools}
\usepackage{mathabx} 
\begin{document} 

\title{Chasing extreme planetary architectures:} 
   \subtitle{I- HD\,196885\,Ab, a super-Jupiter dancing with two stars?}

   \authorrunning{Xtreme team et al.}
   \author{G. Chauvin\inst{1,2}, M. Videla\inst{3,1}, H. Beust\inst{4}, R. Mendez\inst{5}, A.~C.~M. Correia\inst{6,7}, S. Lacour\inst{8}, A. Tokovinin\inst{9}, J. Hagelberg\inst{10}, F. Bouchy\inst{10}, I. Boisse\inst{11}, C. Villegas\inst{3,1}, M. Bonavita\inst{12}, S. Desidera\inst{13}, V. Faramaz\inst{14}, T. Forveille\inst{4}, A. Gallenne\inst{15,1}, X. Haubois\inst{16}, J.S. Jenkins\inst{17,18}, P. Kervella\inst{8}, A.-M. Lagrange\inst{8}, C. Melo\inst{16}, P. Thebault\inst{8}, S. Udry\inst{10} and D. Segransan\inst{10}} 

   \institute{$^{1}$ Unidad Mixta Internacional Franco-Chilena de Astronom\'{i}a, CNRS/INSU UMI 3386 and Departamento de Astronom\'{i}a, Universidad de Chile, Camino El Observatorio 1515, Las Condes, Santiago, Chile\\ 
   $^{2}$ Laboratoire J.-L. Lagrange, Universit\'e Cote d’Azur, CNRS, Observatoire de la Cote d’Azur, 06304 Nice, France\\
   $^{3}$Department of Electrical Engineering, Facultad de Ciencias F\'{i}sicas y Matem\'{a}ticas, Universidad de Chile, Tupper 2007, Santiago, Chile\\
   $^{4}$ Univ. Grenoble Alpes, CNRS, IPAG, F-38000 Grenoble, France \\   
   $^{5}$Department of Astronomy, Facultad de Ciencias F\'{i}sicas y Matem\'{a}ticass, Universidad de Chile, Camino El Observatorio 1515, Las Condes, Santiago, Chile\\
   $^{6}$ CFisUC, Departamento de F\'isica, Universidade de Coimbra, 3004-516 Coimbra, Portugal\\
   $^{7}$ IMCCE, Observatoire de Paris, PSL Universit\'e, 77 Av. Denfert-Rochereau, 75014 Paris, France\\
   $^{8}$ LESIA, Observatoire de Paris, PSL Research University, CNRS, Sorbonne Universités, UPMC Univ. Paris 06, Univ. Paris Diderot, Sorbonne Paris Cité, 5 place Jules Janssen, 92195 Meudon, France\\
   $^{9}$ Cerro Tololo Inter-American Observatory, Casilla 603, La Serena, Chile\\
   $^{10}$ Geneva Observatory, University of Geneva, 51 ch. Pegasi, CH-1290 Versoix, Switzerland \\ 
   $^{11}$ Aix Marseille Univ, CNRS, CNES, LAM, Marseille, France\\
   $^{12}$ School of Physical Sciences, Faculty of Science, Technology, Engineering and Mathematics, The Open University, Walton Hall, Milton Keynes, MK7 6AA\\
   $^{13}$ INAF - Osservatorio Astronomico di Padova, Vicolo dell’ Osservatorio 5, 35122, Padova, Italy\\ 
   $^{14}$ Steward Observatory, Department of Astronomy, University of Arizona, 933 N. Cherry Ave, Tucson, AZ 85721, USA\\
   $^{15}$ Universidad de Concepción, Departamento de Astronomía, Casilla 160-C, Concepción, Chile\\
   $^{16}$ European Southern Observatory, Alonso de Cordova 3107, Casilla 19001, Vitacura, Santiago, Chile \\
   $^{17}$ N\'ucleo de Astronom\'ia, Facultad de Ingenier\'ia y Ciencias, Universidad Diego Portales, Av. Ej\'ercito 441, Santiago, Chile
   $^{18}$ Centro de Astrof\'isica y Tecnolog\'ias Afines (CATA), Casilla 36-D, Santiago, Chile\\
}

   \date{Received July, 2022; accepted September, 2022}

 
  \abstract
   {Planet(s) in binaries are unique architectures for testing predictions of planetary formation and evolution theories in very hostile environments. 
Their presence in such a highly perturbed region poses a
clear challenge from the early phase of planetesimals accretion to the dynamical evolution and stability through a very long period of time (several Gyrs in some case). }
   {The combination of radial velocity, speckle interferometry, high-contrast imaging and high-precision astrometry with interferometry, offers a unique and unprecedented set of observable
   to push the exploration of the extreme planetary system HD\,196885. By dissecting the physical and orbital properties of each component, we aim at shedding light on its global architecture and stability.}
   {We used the IRDIS dual-band imager of SPHERE at VLT, and the speckle interferometric camera HRCAM of SOAR, to acquire high-angular resolution images of HD\,196885\,AB between 2015 and 2020. 
   Radial velocity observations started in 1983 with CORAVEL at OHP have been extended with a continuous monitoring with CORALIE at La Silla, and ELODIE and SOPHIE at OHP over almost 40\,yr extending the radial velocity measurements HD\,196885\,A and resolving both the binary companion and the inner giant planet HD\,196885\,Ab.  Finally, we took advantage of the exquisite astrometric precision of the dual-field mode of VLTI/GRAVITY (down to 30\,$\mu$as) to monitor the relative position of HD\,196885\,A and B to search for the 3.6\,yr astrometric wobble of the circumprimary planet Ab imprinted on the binary separation.}
   {Our observations enable to accurately constrain the orbital properties of the binary HD\,196885\,AB, seen on an inclined and retrograde orbit ($i_{\rm AB}=120.43$\,deg) with a semi-major axis of 19.78\,au, and an eccentricity of 0.417.
 The GRAVITY measurements confirm for the first time the nature of the inner planet HD\,196885\,Ab by rejecting all families of pole-on solutions in the stellar or brown dwarf masses. The most favored island of solutions is associated with a Jupiter-like planet ($M_{\rm{Ab}}=3.39$\,M$_{\rm Jup}$), with moderate eccentricity ($e_{\rm AaAb}=0.44$), and inclination close to $143.04$\,deg. This results points toward a significant mutual inclination ($\Phi=24.36$\,deg) between the orbital planes (relative to the star) of the binary companion B and the planet Ab. Our dynamical simulations indicate that the system is dynamically stable over time. Eccentricity and mutual inclination variations could be expected for moderate von Zipele Kozai Lidov cycles that may affect the inner planet.}
    {}

   \keywords{Instrumentation: adaptive optics, high angular resolution, interferometry -- Methods: observational -- Stars: individual: HD\,196885 -- Planetary systems}

    \titlerunning{Chasing the extreme planetary systems}
   \maketitle

%
\section{Introduction}

Even though binaries were initially banished from exoplanet hunting 
  campaigns, more than 200 planets\footnote{A total of 217 planets in binaries distributed over 154 systems are known on July, 2021 (https://www.univie.ac.at/adg/schwarz/multiple.html)} have been  so far discovered in multiple systems orbiting one component or
  both components, referring to the S-type and P-type configuration respectively (see classification by \citealt{1982OAWMN.191..423D}). These systems have been initially discovered by accident like Gliese 86\,Ab \citep{2000A&A...354...99Q} and $\gamma$ Cep\,Ab \citep{2003ApJ...599.1383H}, but more recently, dedicated surveys in radial velocity (RV), transit and imaging have been targeting them \citep[e.g.][]{2021FrASS...8...16F}. Given that the formation of stars in multiple systems is a frequent byproduct of stellar formation, a natural question is to understand how the presence of a stellar companion can affect the planetary formation process. Planetary formation theories have been restricted for years to the case of a single star environment to understand the formation of our own Solar System \citep{2008ASPC..398..235M}, and more recently investigated for binary stars \citep[e.g.][]{2021A&A...652A.104S}. 

For the most frequent dynamical configuration observed for planet(s) in binaries, the S-type circumprimary one with a planet orbiting one component of the binary (generally the most massive one), models have predicted that the presence of a very close binary companion can hinder planetary formation around a primary star in several ways. The most problematic issue is probably not the disk truncation \citep[see][]{2015ApJ...799..147J}, but rather the dynamical excitation that could prevent mutual planetesimal accretion for objects in the 100\,m-to-100\,km size range \citep{2006Icar..183..193T,2011CeMDA.111...29T,2008MNRAS.386..973P,2011A&A...528A..40F}, hence obstructing the formation of a planet by core accretion, or ejecting the planet in unstable systems \citep[e.g.][]{2015pes..book..309T}. 
Dedicated observing campaigns suggest a transition at typical separation within 50-300\,au indicating that short or intermediate-period binaries have statistically less chance to host planets or brown dwarf companions, compared to wide binaries that on the contrary have no influence on the architectures of planetary systems \citep{2021AJ....162..192Z}. This effect seems to be corroborated by the study of young stars for which short-separation ($\lesssim 100$\,au) binaries have a lower probability of hosting circumstellar dust in the innermost few au around each star, therefore with a depleted reservoir of solids for the formation of planets by core accretion \citep{2010ApJ...709L.114D}.

\begin{figure}[t]
     \centering \includegraphics[width=\columnwidth]{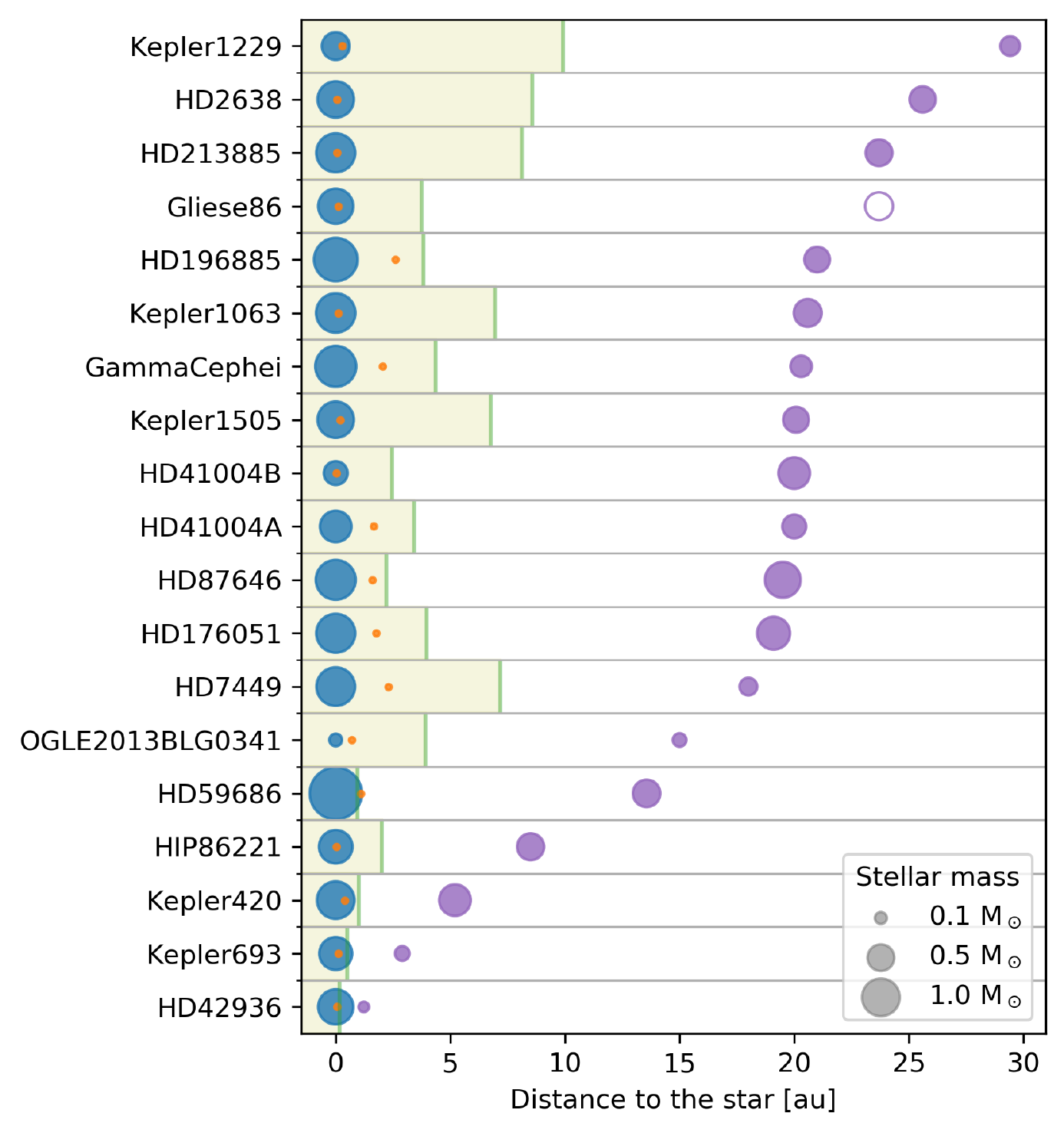}
    \caption{Overview of the binary star systems. The sizes of the circles
representing the primary and secondary star indicate their masses according
to the scale reported on the \textit{Bottom-Right}. The planet masses are not to scale. 
The green zone marks the region of orbital stability for a coplanar configuration in the context of an S-type configuration.} 
    \label{fig:xtremes}
\end{figure}

In the last decade, more intensive and dedicated surveys have been dedicated to search for S-type and P-type\footnote{P-type orbits occur when the planet orbits both binary components, and generally referred as circumbinary planet} exoplanetary systems using (and often combining) RV, transit, high-contrast and direct imaging observations. RV surveys, able to tackle multiple stellar signals at the precision required for detecting planets, were specifically carried out \citep{2005AAS...207.8402K,2007arXiv0705.3173E,2009PASJ...61...19T,2010ASSL..366..105D}. \cite{2002ApJ...568L.113Z} initially showed that the planet properties in binaries were different to planets around single stars. \cite{2004ASPC..321...93E} confirmed that the most massive short period planets were found in binary systems and that planets with short ($P\lesssim40$~days) periods in binaries were likely to have low eccentricities. \cite{2008A&A...480L..33T} finally found that extremely high eccentric planets were all in binaries. 

In parallel to Doppler surveys, high-contrast imaging campaigns have been able to access intermediate-separation ($20-100$~au) binaries where the companion influence is the most expected. Several deep imaging studies have revealed a number of additional close companions to known exoplanetary
hosts \citep{2002ApJ...581..654P,2002ApJ...566.1132L,2006A&A...456.1165C,2007arXiv0705.3173E,2014ApJ...791...35L,2015MNRAS.450.3127M,2016AJ....152...18B,2021A&A...649A.156G}. \cite{2007A&A...474..273E} carried out a dedicated survey to explore the multiplicity in two samples of stars with and without planets detected in RV in order to test the impact of duplicity on the giant planet occurrence. They found a lower corrected binary fraction for the planet-host sub-sample, particularly for physical separations shorter than $100$~au \citep{2008ASPC..398..179E}.

For longer period ($\gtrsim100$~au) binaries, catalogs compiled by \cite{2006ApJ...646..523R}, \cite{2007A&A...462..345D}, and \cite{2007A&A...468..721B} further updated by \cite{2020Galax...8...16B} studied the effect of duplicity, and did not find any significant discrepancies in the planet frequency between binary and single stellar systems. More recently, the use of the Gaia offered the opportunity of a systematic search for (sub)-stellar companions to exoplanet host \citep{2019A&A...623A..72K,2019MNRAS.490.5088M,2021FrASS...8...14M}.
\cite{2021FrASS...8...16F} recently derived an overall raw multiplicity rate of 23.2$\pm$1.6$\%$ for hosts to exoplanets and brown dwarf companions, considering a volume-limited sample up to 200\,pc cross-matched with the Gaia Data Release 2 catalog 
\citep{2016A&A...595A...2G,2018A&A...616A...1G}.

A complementary approach has been to target individual systems, and particularly the extreme ones that provide theoreticians with an ideal test bed for the theories of planetary formation, evolution and dynamical stability in a very hostile environment. Today, there are about 16 extreme exoplanetary systems (binaries with small semi-major axes of less than about 20~au hosting a planet; see Fig.\,\ref{fig:xtremes}). One emblematic case is $\gamma$ Cep for which \cite{2018RNAAS...2....7B} managed to derive the mass and the orbital  properties of the circumprimary planet using accurate astrometric measurements with \textit{HST}. Orbiting at 1.94\,au in a highly mutually inclined orbit ($\Phi=70\pm13$\,deg), the formation and survival of this $9.4^{+0.7}_{-1.1}$\,M$_{\rm Jup}$ exoplanet remains unresolved \cite[e.g.][]{2017MNRAS.466.1555B}.  Another interesting system is HD\,196885 that \cite{2006A&A...456.1165C} discovered in the course of a VLT/CFHT deep imaging survey of exoplanet hosts. In this paper, we present the results of the long-term monitoring of this system combining RV, direct imaging, speckle and dual-field interferometry. The physical properties of the HD\,196885 exoplanetary system are reported in Section\,2. The observations and data reduction are presented in Section\,3, together with the results in Section\,4. In Section\,5, we finally discuss their implication in terms of system architecture, formation and stability.    

\section{The HD\,196885 exoplanetary system}

\begin{table}[t]
   \centering
    \small
    \caption{Physical properties of the extreme planetary system HD\,196885, the stellar host (Aa), the binary companion (B) and the circumprimary planet (Ab) based on the latest studies from \cite{2008A&A...479..271C,2009ApJ...703.1545F,chauvin2011,2021A&A...649A...1G}. We consider below that the system is composed of the inner binary subsystem AaAb (composed by the pri-
mary star Aa and the exoplanet Ab), and the outer binary subsystem AB (composed by the center of
mass of the inner orbit A and the companion star B)}
    \begin{tabular}{|l|l|}
        \multicolumn{2}{c}{Primary star: HD\,196885\,Aa}  \\
         \hline
         Spectral type  \hspace{4cm}& F8V        \hspace{2cm}          \\
         G (mag)        & $6.276973\pm0.002767$  \\
         V (mag)        & $6.385\pm0.010$  \\
         H (mag)        & $5.190\pm0.023$  \\
         Teff (K)        & $6340\pm39$  \\
         log(g)  (dex)        & $4.46\pm0.02$       \\
         Fe/H  (dex)        & $0.29\pm0.05$       \\
         Distance (pc)      & $34.00\pm0.04$    \\
         $\mu_{\alpha}$ (mas/yr) & $71.915\pm0.032$ \\
         $\mu_{\delta}$ (mas/yr) & $89.318\pm0.026$ \\
         Age (Gyr)            & $1.5-3.5$  \\
         M$_{\rm Aa}$ (M$_\odot$)         & $1.3\pm0.1$     \\
         log(L/L$_\odot$)     & $2.4\pm0.1$    \\
         \hline\noalign{\medskip} 
         \multicolumn{2}{c}{Exoplanet: HD\,196885\,Ab}          \\ 
         \hline
         M$_{\rm{Ab}}$sin$(i_{\rm{AaAb}})$ ($M_{\rm Jup}$)   & $2.98\pm0.05$     \\
         $P_{\rm AaAb}$ (yr)   & $3.63\pm0.01$                           \\
         $e_{\rm AaAb}$               & $0.48\pm0.02$                          \\
         $\omega_{\rm AaAb}$ (deg)               & $93.2\pm3.0$                          \\
         $t_{\rm P AaAb}$               & $2002.85\pm0.02$                          \\
         $a_{\rm AaAb}$ (au)              & $2.6\pm0.1$                          \\
         \hline\noalign{\medskip} 
         \multicolumn{2}{c}{Binary companion: HD\,196885\,B}  \\
         \hline
         Spectral type   & M$1\pm1$                   \\
         K$_{\rm B}$ (mag)        & $6.398\pm250$  \\
         M$_{\rm B}$  (M$_\odot$)         & $0.45\pm0.01$     \\
         $P_{\rm AB}$ (yr)   & $72.06\pm4.59$                           \\
         $e_{\rm AB}$               & 0.$42\pm0.03$                          \\
         $\omega_{\rm AB}$  (deg)             & $-118.1\pm3.1$                          \\
         $\Omega_{\rm AB}$ (deg)              & $79.8\pm0.1$                          \\
         $i_{\rm AB}$ (deg)               & $116.8\pm0.7$                          \\
         $t_{\rm P AB}$               & $1985.59\pm0.39$                          \\
         $a_{\rm AB}$ (au)              & $21.00\pm0.86$                          \\
         \hline \noalign{\smallskip}\noalign{\smallskip}        
             \end{tabular} \quad
    \label{tab:parasys}
\end{table}

HD\,196885 is an F8V ($V = 6.385$\,mag, $H = 5.19$\,mag) star located at $34.00\pm0.04$\,pc (Gaia EDR3, \citealt{2021A&A...649A...1G}). Based on CORALIE spectra, Sousa et al. (2006) derived a spectroscopic temperature, surface gravity, and metallicity of T$_{\rm eff} = 6340\pm39$\,K, log(g)$ = 4.46\pm0.02$ and [Fe/H]$= 0.29\pm0.05$, respectively. The projected stellar rotational velocity $v$.sin($i$) was estimated to $7.3\pm1.5$\,km/s from ELODIE spectra \citep{2008A&A...479..271C}. These results were supported by independent Lick observations reported by \cite{2009ApJ...703.1545F}. The chromospheric activity level of log$(R'_{HK}) = -5.01$ \citep{2004ApJS..152..261W} is relatively low when placed into context with the activity distribution of the solar neighborhood \citep{2011A&A...531A...8J}. Bolometric luminosity correction and evolutionary model predictions lead to an estimate of the luminosity and the mass of 2.4\,L$_{\odot}$ and 1.3\,M$_{\odot}$, respectively. The corresponding stellar age derived from evolutionary tracks and from the activity level varies between 1.5 to 3.5 Gyr \citep[see][]{2008A&A...479..271C,2009ApJ...703.1545F}. 
Finally, we looked at wider physical companions within 3 arcmin from HD\,196895 in Gaia EDR3 and found none. The companion BUP\,1908 (192\,arcseconds separation, position angle 8$^{\circ}$, $V=10.4$\,mag, TYC\,1096-637-1) listed in the double star catalog, CCDM, is optical according to Gaia. As expected, HD\,196885 shows a huge signature of binarity (with an S/N of 221) in the Hipparcos-Gaia proper motion anomaly catalog \citep{2022A&A...657A...7K}, but less constraining that the set of observations presented below.

\begin{figure}[t]
     \centering \includegraphics[width=\columnwidth]{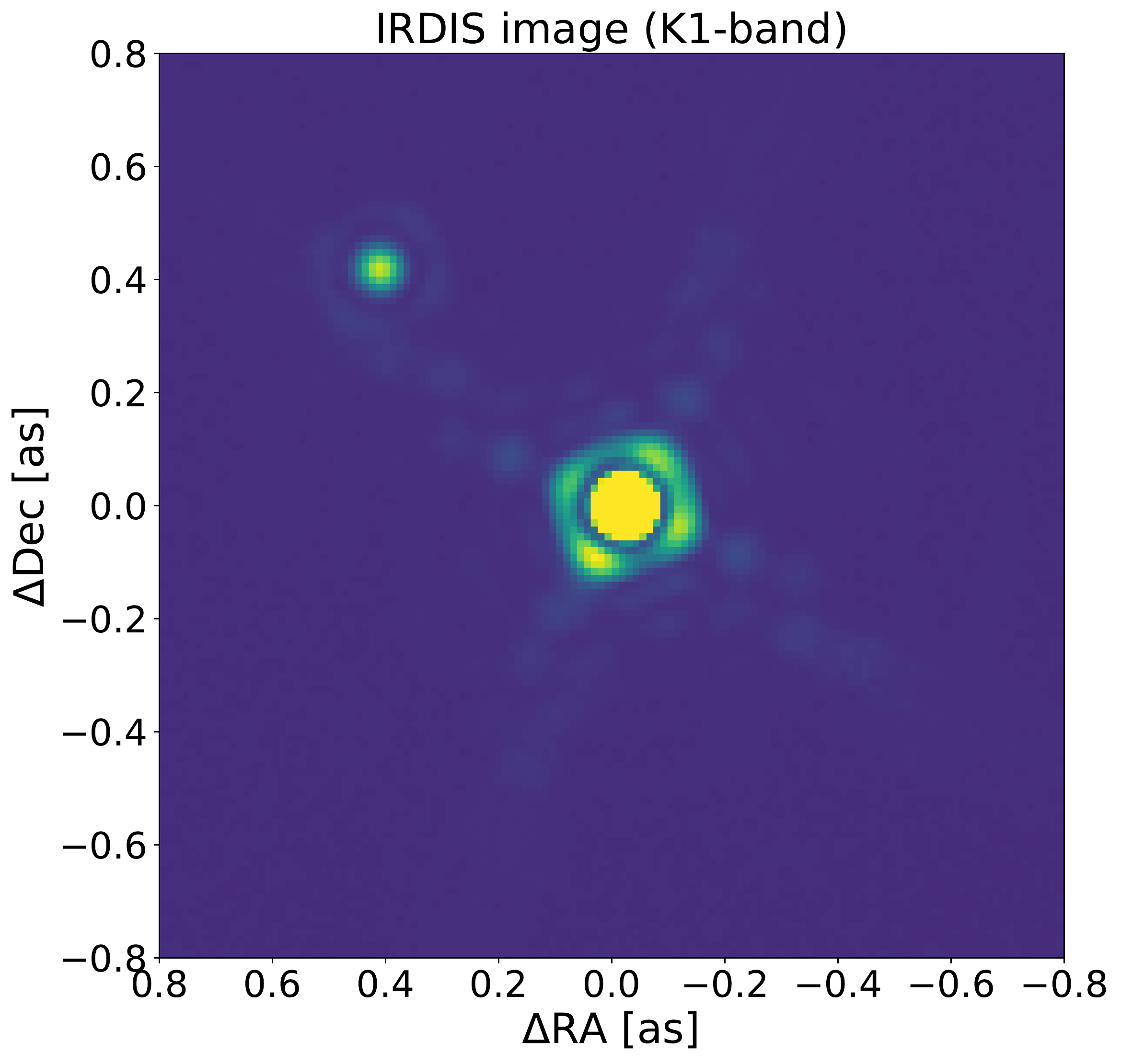}
    \caption{SPHERE/IRDIS observations of HD\,196885\,AB obtained in September, 2017. The early-M1V dwarf companion HD\,196885\,B is resolved at K1-band at a separation of 604\,mas and a position angle of $45.7$\,deg.} 
    \label{fig:irdis}
\end{figure}

In 2004, an RV variation indicating a planetary candidate was measured at Lick observatory and adjusted with a preliminary orbital period of $P = 0.95$\,yr. The result was temporarily reported on the California Planet Search Exoplanet Web site, but finally withdrawn owing a significant residual drift in the orbital solution. Nevertheless, this star was included in a VLT/CFHT deep-imaging survey of stars hosting planets detected by RV observations \citep{2006A&A...456.1165C}. This led them to the discovery of a close ($\sim 20$~au) stellar binary companion to HD\,196885, motivating dedicated follow-up observations with Adaptive-Optics (AO) imaging. The recent image of the system obtained by SPHERE at VLT is shown in Fig.\,\ref{fig:irdis}. Following the CFHT discovery, astrometric monitoring and long-slit spectroscopic observations at VLT with NaCo enabled to resolve the orbital motion of the stellar companion and to confirm its low stellar-mass nature as an M1$\pm$1V dwarf companion located at only $0.7$\,as (23\,au in projected physical separation) and likely to be responsible for the trend seen in the Lick RV residuals \citep{2007A&A...475..723C}. Using a double-Keplerian model for the binary star and the planet to adjust their ELODIE, CORALIE, and CORAVEL observations spread over 14\,yr, \cite{2008A&A...479..271C} confirmed the existence of the planet HD\,196885\,Ab orbiting the primary. They derived a first range of orbital solutions with a minimum mass of M$_{\rm Ab}.$sin$(i_{\rm AaAb}) = 2.96$ M$_{\rm Jup}$, a period of $P_{\rm AaAb} = 3.69\pm 0.03$\,yr, and an eccentricity of $e_{\rm AaAb} = 0.462 \pm 0.026$. Moreover, they found additional constraints for the binary companion HD\,196885\,B with a period of $P_{\rm AB} \ge 40$\,yr, a semi-major axis $a_{\rm AB} \ge 14$\,au, and a minimum mass of M$_{\rm B}$.sin$(i_{\rm AB}) \ge 0.28$\,M$_{\odot}$. Based on Lick observations, \cite{2009ApJ...703.1545F} derived consistent results for both the inner planet and the binary companion. 

In 2011, \cite{chauvin2011} reported a four years astrometric monitoring of the separation between HD\,196885\,A and B using NaCo at VLT. Combined with RV observations, these new measurements enabled them to refine the constraints on the orbital properties of the stellar companion, and to derive an inclination of $i_{\rm AB}=116.8\pm0.7$\,deg, an eccentricity of $e_{\rm AB}=0.42\pm0.03$, a semi-major axis of $a_{\rm AB}=21.00\pm0.86$\,au, and a true mass of of $M_{\rm B}=0.45\pm0.01$\,M$_{\oplus}$,\,consistent with the spectral type M1$\pm$1V derived by \cite{2007A&A...475..723C}.

\section{Observations and data reduction\label{sec:obs}}

\begin{table*}[t]
   \centering
    \small
         \caption{Observation log of the new SPHERE, SOAR, and GRAVITY observations. The exposure time is given by Nexp, the number of exposures, multiplied by NDIT and DIT, which are the number and duration of integrations per exposure (in the case of HRCAM in ms). $\omega$ and $\tau_0$ are the seeing and the coherence time respectively. For the modes, DBI refers to Dual-Band Imaging, SI to Speckle Interferometry and MED for medium spectral resolution interferometry.}
\begin{tabular}{ccccccccccc} 
\hline\hline\noalign{\smallskip}\footnotesize
UT-Date & MJD & Telescope & Instr. & Filter & Mode & Nb Exp. & Nb Integration & Exp. Time & $\omega$ & $\tau_0$ \\   
     &     &   & & &       & & & (s)                         & (")    & (ms)  \\
\noalign{\smallskip}\hline\noalign{\smallskip} 
27-09-2000  & 51826.186 & SOAR & HRCAM     & I&SI   & 3&400&0.024  &  &   \\
27-09-2002  & 52459.690 & SOAR & HRCAM     & I&SI   & 3&400&0.024  &  &   \\
05-06-2015  & 57178.378  & VLT/UT3 & IRDIS    & K1&DBI & 21& 35 & 2  & 1.75 & 1.2  \\
27-09-2015  & 57291.493 & SOAR & HRCAM     & I&SI   & 3&400&0.024  &  &   \\
30-09-2017  & 58026.077 & VLT/UT3 & IRDIS    & K1&DBI & 2&30&2  & 0.72 & 3.4  \\ 
27-05-2018  & 58265.219 & SOAR & HRCAM     & I&SI   & 2&400&0.024  &  &   \\
16-07-2019  & 58680.881 & SOAR & HRCAM     & I&SI   & 2&400&0.024  &  &   \\  
15-08-2019  & 58710.188  & VLTI/UTs & GRAVITY   & K&MED & 3&20&10 & 0.95 & 2.9\\ 
29-05-2021  & 59363.331 & VLTI/UTs & GRAVITY   & K&MED & 3&12&30  & 0.64 & 5.0  \\  
24-08-2021  & 59450.195  & VLTI/UTs & GRAVITY   & K&MED & 3&12&30  & 0.90 & 2.6  \\  
02-10-2021  & 59489.151 & SOAR & HRCAM     & I&SI   & 3&400&0.024  &  &  \\ 
20-10-2021  & 59507.082  & VLTI/UTs & GRAVITY   & K&MED & 3&12&30  & 1.00  & 2.8  \\  
\noalign{\smallskip}\hline\noalign{\smallskip}     
\end{tabular}
\label{tab:obslog}
\end{table*} 

\subsection{Radial velocities}

In an attempt to revisit the physical and orbital properties of the HD 196885 system, we considered all data from past radial velocity surveys already published in \cite{chauvin2011} for the primary host HD\,196885\,A: CORAVEL (9 RV measurements from June 1982 to August 1997), ELODIE (69 measurements from June 1997 to August 2006), CORALIE (33 measurements from April 1999 to November 2002), and Lick (75 measurements from 1998 to 2008). To this, we have added new SOPHIE observations (52 measurements from 2007 to July 2017) to gather a total of 238 radial velocity measurements spanning almost 40\,yr of observation (see \citealt{2008A&A...479..271C,2009ApJ...703.1545F,chauvin2011}, and this work). SOPHIE data were acquired in High-resolution mode as part of the program dedicated to the follow-up of ELODIE candidates. Data from 2007 to June 2011 where reduced thanks to the SOPHIE pipeline \citep{2009A&A...505..853B}. In June 2011, an update of the instrument was done. It consists mainly in the addition of octogonal fibers to light guiding from the telescope to the spectrograph \citep{2013A&A...549A..49B}. Data after this date were then reduced as in \cite{2022A&A...658A.176H}.
  
\begin{table}[t] 
\centering
\small
\caption{Relative astrometry of HD\,196885 B relative to A extracted from NaCo (see \citealt{chauvin2011}),
SOAR, IRDIS, and GRAVITY (refered as GRAV. in the Table) observations. For GRAVITY, the Pearson’s coefficient $\rho$ quantify the correlation between the $\Delta\alpha$ and $\Delta\delta$ uncertainties.}
        \begin{tabular}{llllllll}     
\hline\hline\noalign{\smallskip}\footnotesize
Inst. &MJD     & $\Delta\alpha$   & $\Delta\delta$    &    $\sigma_{\Delta\alpha}$      & $\sigma_{\Delta\alpha}$    &   $\rho$ \\
   &          & (mas)            & (mas)             &    (mas)         & (mas)          &                 \\
\hline\noalign{\smallskip}
SOAR & 51826.186  & 650.0 & 135.8 & 5.0 & 5.0 & \\
SOAR & 52459.690  & 709.7 & 143.1 & 5.0 & 5.0 & \\
\noalign{\smallskip}
NaCo & 53583.238   & 658.5 & 273.6 & 1.7 & 1.6 &  \\
NaCo & 53974.086   & 649.8 & 292.6 & 1.9 & 2.4 & \\
NaCo & 54337.053   & 640.2 & 309.7 & 3.1 & 3.2 & \\
NaCo & 54645.259   & 630.9 & 326.3 & 3.1 & 3.2 & \\
NaCo & 55070.108   & 614.5 & 342.1 & 3.0 & 2.9 & \\
\noalign{\smallskip}
IRDIS & 57178.378   & 497.0 & 405.0 & 2.0 & 2.0 & \\
SOAR & 57291.493  & 491.0 & 407.7 & 3.0 & 3.0 & \\
IRDIS & 58026.077   & 433.0 & 422.0 & 2.0 & 2.0 & \\
SOAR & 58265.219  & 417.8 & 420.7 &3.0 &3.0 & \\
SOAR & 58680.881  & 383.1 & 420.1 &3.0 &3.0 &  \\
GRAV. &58710.188    & 377.115 & 425.575 & 0.017 & 0.037 & -0.61 \\
GRAV. &59363.331    & 320.243 & 428.407 & 0.015 & 0.043 & -0.74 \\
GRAV. &59450.195 & 312.573 & 428.333 & 0.034 &0.049 & -0.96\\
SOAR & 59489.151 & 385.2 & 3.0 & 3.0 & 3.0 & \\
GRAV. & 59507.082 & 307.471 & 428.540 & 0.038 & 0.037 & -0.65 \\ 
\noalign{\smallskip}\hline\noalign{\smallskip}     
\end{tabular}
\label{tab:astrometry}
\end{table} 

\subsection{SPHERE adaptive-optics imaging}

Following previous astrometric observations at VLT obtained with the NaCo instrument, the HD\,$196885$ system has been monitored during the SpHere INfrared survey for Exoplanets \citep[SHINE, ][]{Chauvin2017_shine,Desidera2021_shine_paperI_sample_definition,Langlois2021} at $2$ different epochs on 5 June 2015 and 30 September 2017 (see Table~\ref{tab:obslog}) using the VLT/SPHERE high-contrast instrument \citep{Beuzit2019_sphere}. The observations were obtained with the modes IRDIFS-EXT that combine simultaneously the IRDIS \citep{Dohlen2008} and IFS instruments \citep{Claudi2008}, although IRDIS observations were mainly used in the current study. The IRDIFS-EXT mode combines IRDIS in dual-band imaging \citep[DBI][]{Vigan2010dbi} mode with the K1K2 filter doublet  
$\lambda_{K1}=2.103\pm0.102~\mathrm{\mu m}$, $\lambda_{K2}=2.255\pm0.109~\mathrm{\mu m}$, and IFS in the YJH ($0.97$--$1.66\,\mu$m) setting. Each observing sequence was performed with the field-tracking mode. The detail about the observations are reported in Table~\ref{tab:obslog}. Only the IRDIS observations at K1-band (with a lower background noise) were used to extract the relative astrometry of HD\,196885\,B relative to A.

\subsection{SOAR speckle interferometry}

Complementary astrometric measurements of HD\,$196885$ have been obtained with the High-Resolution Camera (HRCAM) \cite{2021AJ....162...41T} at the SOAR 4.1-m telescope \cite{tokovinin2008}\footnote{For up-to-date details of the instrument see \url{http://www.ctio.noirlab.edu/~atokovin/speckle/}}. While regular observations are made in the Stroemgren $y$, Cousins $I$, or $H \alpha$ filters, all observations of HD~196885 were done with the $I$ filter with a resolution of 36\,mas.
The data have been processed by a custom IDL pipeline developed by A. Tokovinin to extract the binary parameters (separation, position angle, and magnitude difference) by fitting the secondary peaks of the auto-correlation functions or the fringes from the power spectrum of the observations obtained by Fourier transform as explained in \cite{TokMasHar2010}. HRCAM routinely delivers a precision of 1-3~mas in angular separation for objects brighter than $V \sim 12$. On every observing night "calibration binaries" are included, these are binaries with very well known orbits (grade 5 on the USNO orbit catalogue), from which the measurements are calibrated, leading to systematic errors less than 0.1\,deg in position angle, and better than 0.2\% in scale, i.e., smaller than the internal precision. The final precision of these measurements depends on a number of factors, but in this paper we adopt a uniform uncertainty of 3~mas, which is representative of all the measurements \citep{Toko2018}. The Observing log and the astrometric results are reported in Tables~\ref{tab:obslog} and \ref{tab:astrometry}. 

\begin{figure*}[t]
     \centering 
     \includegraphics[width=\textwidth]{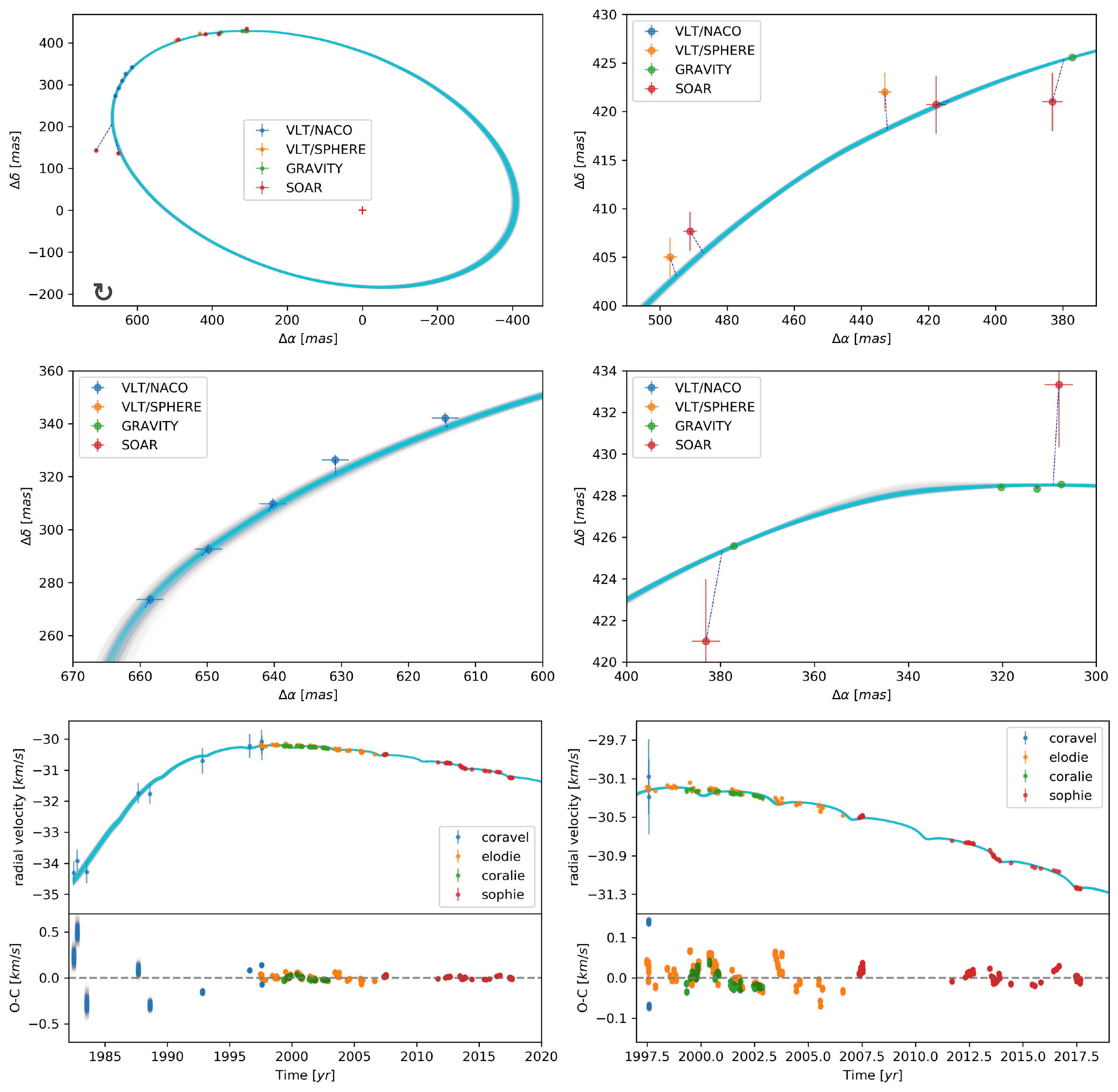}
    \caption{\textit{Top-Left}: Astrometric position of HD\,196885\,B relative to A combining SOAR, NaCo, SPHERE and GRAVITY observations from August 2005 to August, 2021. Specific zoomed plots on the NaCo astrometric measurements obtained between August, 2005 and August, 2009 (\textit{Middle-Center}), the SPHERE and SOAR ones from June, 2015 to May, 2018 (\textit{Top-Right}), and the GRAVITY ones from May, 2019 to August, 2021  (\textit{Middle-Right}) are shown. \textit{Bottom-Left}: Radial velocity measurements of HD\,196885\,A from CORAVEL, ELODIE, CORALIE, Lick and SOPHIE covering 40\,yr of monitoring. \textit{Top-Right}: Zoom on the radial velocity signal induced by HD\,196885\,Ab. The results of the orbital fitting exploration are reported with the \textit{thick blue} curves.} 
    \label{fig:orbitfit}
\end{figure*}

\subsection{VLTI/GRAVITY dual-field interferometry}

HD\,$196885$ was also monitored in the context of monitoring program with GRAVITY \citep{2017A&A...602A..94G} and the UTs in MED resolution at K-band. Over two years, we have obtained a total of four observations on 15 August 2019, 29 May 2021, 25 August 2021, and 25 September 2021. The weather conditions ranged from good to very good, and the observing log of the observations is also reported in Table\,\ref{tab:obslog}.

These observations were obtained with the astrometric dual-field mode in off-axis: in the first exposure, the fringe tracker single-mode fiber is centered on the primary star HD\,196885 A while the scientific single-mode fiber is centered on the stellar companion HD\,196885\,B. In a second exposure, the two fibers are swapped, to calibrate the metrology offset. This observation mode is different from the dual-field mode in on-axis, where the planet is bright enough to be directly observed \citep{2019A&A...623L..11G}. Here, the GRAVITY interferometer only observed the A and B components. The observations were pre-processed with the Public Release 1.5.0 (1 July 2021) of the ESO Gravity pipeline \citep{2014SPIE.9146E..2DL}. The reduced data are then processed to extract astrometry using the GRAVITY python tools offered to the community \footnote{\url{https://version-lesia.obspm.fr:/repos/DRS_gravity/gravi_tools3}}. The astrometric values are presented in Table\,\ref{tab:astrometry}.

\section{Resolving the astrometric wobble due to the giant planet HD\,196885\,Ab\label{sec:orbits}}

The main objective of our GRAVITY program was to measure the astrometric wobble in the separation of a binary system related to presence of the giant planet, and in such a circumstance be able to derive the orbital properties of the binary companion and the giant planets, but also to confirm the planetary nature of HD\,196885\,Ab. To date, such observation have only been performed in the context of the $\gamma$ Cep system by \cite{2018RNAAS...2....7B}.  Combining all the available measurements obtained from radial velocity, adaptive-optics imaging, speckle interferometry, and dual-field interferometry, we have revisited the system properties using a Bayes-based orbital elements estimation inspired in a previous study by our team to triple hierarchical stellar systems \citep{2021PASP..133g4501V}.

\subsection{Code description and set up}

The orbital parameters estimation code has been developed in a Bayesian Markov-Chain Monte Carlo-based framework, in the context of the so-called hierarchical approximation to combine astrometric with spectroscopic measurements for multiple systems.  
The current implementation of this code\textbf{\footnote{The original software is available on GitHub: \texttt{BinaryStars} codebase: \url{https://github.com/mvidela31/BinaryStars} under a 3-Clause BSD License. Its evolution used in this study will soon be released.}} is an adaptation of the Bayesian inference algorithm for binary stars developed by \citet{videla2022bayesian}, which is based on the No-U-Turn Sampler (NUTS) algorithm \citep{hoffman2014no}. The NUTS algorithm is an MCMC method that avoids the random-walk behavior and the sensitivity to correlated parameters of other MCMC methods commonly used in astronomy (such as the Metropolis-Hastings within Gibbs sampler \citep{ford2005quantifying}, the Parallel Tempering sampler \citep{gregory2005bayesian,gregory2011bayesian}, the Affine Invariant MCMC Ensemble sampler \citep{hou2012affine}, and the Differential Evolution Markov Chain sampler \citep{nelson2013run}) by incorporating first-order gradient information of the parameters space to guide the sampling steps with an adaptive criterion for determining their lengths. The NUTS algorithm qualities allow for performing a quick and efficient estimation of the target distribution, specially in the case of the high-dimensionality problem in hand. 
We estimate the posterior distribution of the joint orbital parameter space of a triple hierarchical system that we adapted to explore the planetary architecture in S-type binaries. The parameter space consists of the orbital parameters of the inner binary subsystem AaAb (composed by the primary star Aa and the exoplanet Ab), and the orbital parameters of the outer binary subsystem AB (composed by the center of mass of the inner orbit A and the companion star B), assuming of course the same value of the parallax $\pi$ for both subsystems. Therefore, the parameter space is defined by the vector $\vartheta=\{P_{\rm AaAb},t_{\rm P AaAb},e_{\rm AaAb},a_{\rm AaAb},\omega_{\rm AaAb},\Omega_{\rm AaAb},i_{\rm AaAb},q_{\rm AaAb}\}\cup\{P_{\rm AB},t_{\rm P AB},e_{\rm AB},\omega_{\rm AB},\Omega_{\rm AB},i_{\rm AB},q_{\rm AB},\pi,V_0\}$\footnote{The semi-major axis of the outer orbit $a_{\rm AB}$ is not part of the parameter space to be sampled since it is deterministically computed using the equivalence between the total mass of the inner system AaAb and the mass of the primary object of the outer system AB, i.e., $M_{\rm Aa}+M_{\rm Ab}=M_{\rm A}$.}. We use uniform priors on all the orbital parameters according to its physical valid range (e.g., $q_{\rm AaAb}\sim \mathcal{U}_{[0,1]}$), except on the system parallax, where we use a Gaussian prior $\pi\sim\mathcal{N}(29.408,0.027^2)$\,mas (according to the value and standard deviation reported by \textit{Gaia} DR3, \citealt{2016A&A...595A...1G}). Based on the spectral types reported for the primary and secondary stellar components reported in Table~\ref{tab:parasys}, we also used truncated Gaussian priors on the stellar masses $M_{\rm Aa}\sim\mathcal{N}(1.245, 0.067^2)$\,$\rm{M}_\odot$ and $M_{\rm B}\sim\mathcal{N}(0.485, 0.059^2)$\,$\rm{M}_\odot$ within the interval $[\mu-2\sigma,\mu+2\sigma]$ (since it accounts $\sim 95\%$ of the Gaussian probability density function), using the recent calibration provided by \citet{Abuset20}\footnote{Note that the masses $M_{\rm Aa}$, $M_{\rm Ab}$ and $M_{\rm B}$ are not in the set of parameters to sample since they are generated from $\vartheta$.}. We run 10000 iterations on each of four independent Markov chains, discarding the first half for the warm-up phase. Each chain was initialized from an starting point determined through the quasi-Newton method L-BFGS \citep{liu1989limited}. The code was implemented using the probabilistic programming language Stan \citep{carpenter2017stan}.

\begin{figure*}[t]
    \centering \includegraphics[width=\textwidth]{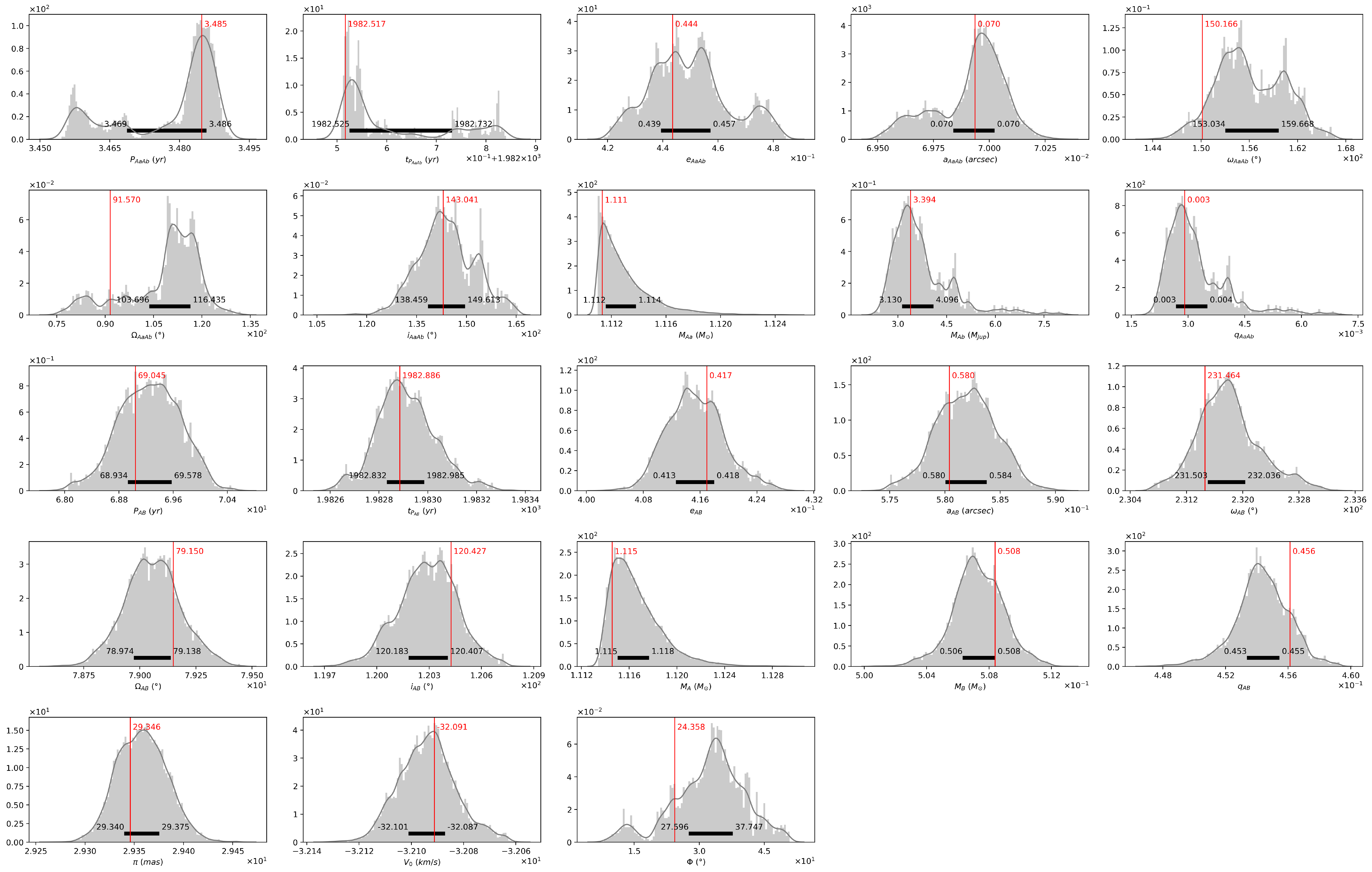}
    \caption{Marginal posterior distributions of the physical and orbital parameters of HD\,196885\,Ab including the period, periastron passage, eccentricity, the semi-major axis...  (two first rows). On the two last rows are reported the marginal posterior distributions for HD\,196885\,B. The maximum a posteriori (the most probable sample of the posterior distribution) are reported in \textit{red}.} 
    \label{fig:params}
\end{figure*}
\subsection{A massive Jupiter on a mutually inclined orbit relative to the stellar companion?}

The results of the orbital fitting exploration led to the identification of nine for families of solutions (considering a truncated Gaussian prior for the mass of HD 196885 A) that are reported in Table\,\ref{tab:sol_trunc}. Their log-posterior density (up to a constant) distributions is also shown in Fig\,\ref{fig:logpost_trunc}. In Table\,\ref{tab:params} and in Fig.~\ref{fig:orbitfit} (in \textit{blue}), we report the most likely solutions of the family Sol.\,2 which are compared with the observed relative astrometric and radial velocity measurements of the system. The posterior solutions of each orbital and physical parameters of the exoplanetary companion HD\,196885\,Ab, and the ones of the stellar companion B are shown in Fig.\,\ref{fig:params}. 

The long-term monitoring over almost 40\,yr combined with 15\,yr of direct imaging set a relatively tight constraints on the orbital properties of the stellar companion. The solutions are very similar to the ones already derived by \citet{chauvin2011} indicating a moderately eccentric ($e_{\rm{AB}}=0.417$) orbit, a 70\,yr period, an inclination of $120.43$\,deg, and a time to periastron in 1982.87\,yr. However, the addition of SOAR, SPHERE, and GRAVITY astrometric measurements from the last decade enables us to nail down previous uncertainties. 

For the planet HD\,196885\,Ab, the gain of almost two orders of magnitude in astrometric accuracy obtained by GRAVITY at VLTI ($30-50$\,$\mu$as precision) compared to what was regularly achieved by NaCo and SPHERE at VLT ($1-2$\,mas precision) completely opens a new window to detect and characterize the periodic wobble caused by the S-type planet on the angular separation between both stellar components A and B. The four GRAVITY measurements obtained in August 2019, May 2021, August 2021, and October 2021 covering less than 60\% of the orbital period of HD\,196885\,Ab, allow to unambiguously reject all families of solutions associated with stellar and brown dwarf masses for Ab, and to pinpoint one specific family of retrograde solutions of Jupiter-like planets with low masses ($M_{\rm{Ab}}=3.39$\,M$_{\rm Jup}$), with eccentricity ($e_{\rm{AaAb}}=0.44$), and inclination close to $143.04$\,deg with a moderate mutual inclination with B ($\Phi=24.36$\,deg). 

We note that other islands of solutions remain possible in the planetary mass regime but for different orbital configurations: retrograde and prograde solutions with higher mutual inclination ($\Phi=60-80$\,deg or $\Phi=110-120$\,deg, respectively) for low masses for Ab ($M_{\rm Ab}=2-4$\,M$_{\rm Jup}$) and moderate eccentricities, as well as one family of prograde solution for higher masses ($M_{\rm Ab}=12-14$\,M$_{\rm Jup}$), moderate eccentricity and mutual inclination ($\Phi=130$\,deg).


\begin{table}[h]
    \caption{Summary of the orbital parameters derived for HD\,196885\,Ab and HD\,196885\,B: the semi-major axis ($a$), the eccentricity ($e$), the inclination ($i$), the longitude of ascending node ($\Omega$), the argument of periastron ($\omega$), the time of periastron passage ($t_P$). We report the maximum a posteriori estimates (the most probable solution), with the first and third quartiles (as an uncertainty measure) as subscripts and superscripts, respectively. Note that the MAP values are not necessarily equal to the second quartile (the median).}    
    \centering \small
    \input{summary_table}
    \label{tab:params}
\end{table}
\begin{figure*}[t]
  \makebox[\textwidth]{
    \includegraphics[width=0.3\textwidth]{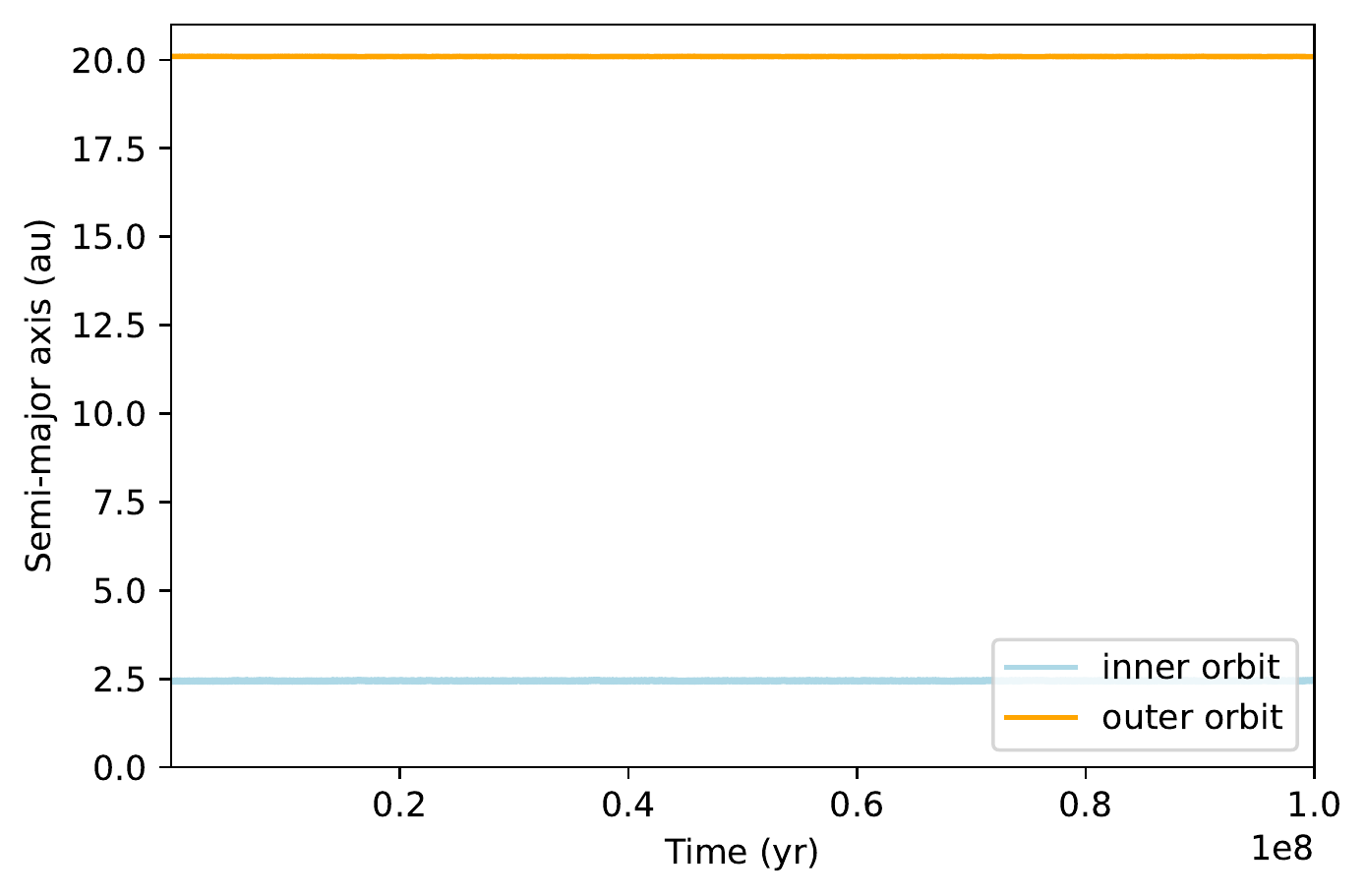}
    \hfil
    \includegraphics[width=0.3\textwidth]{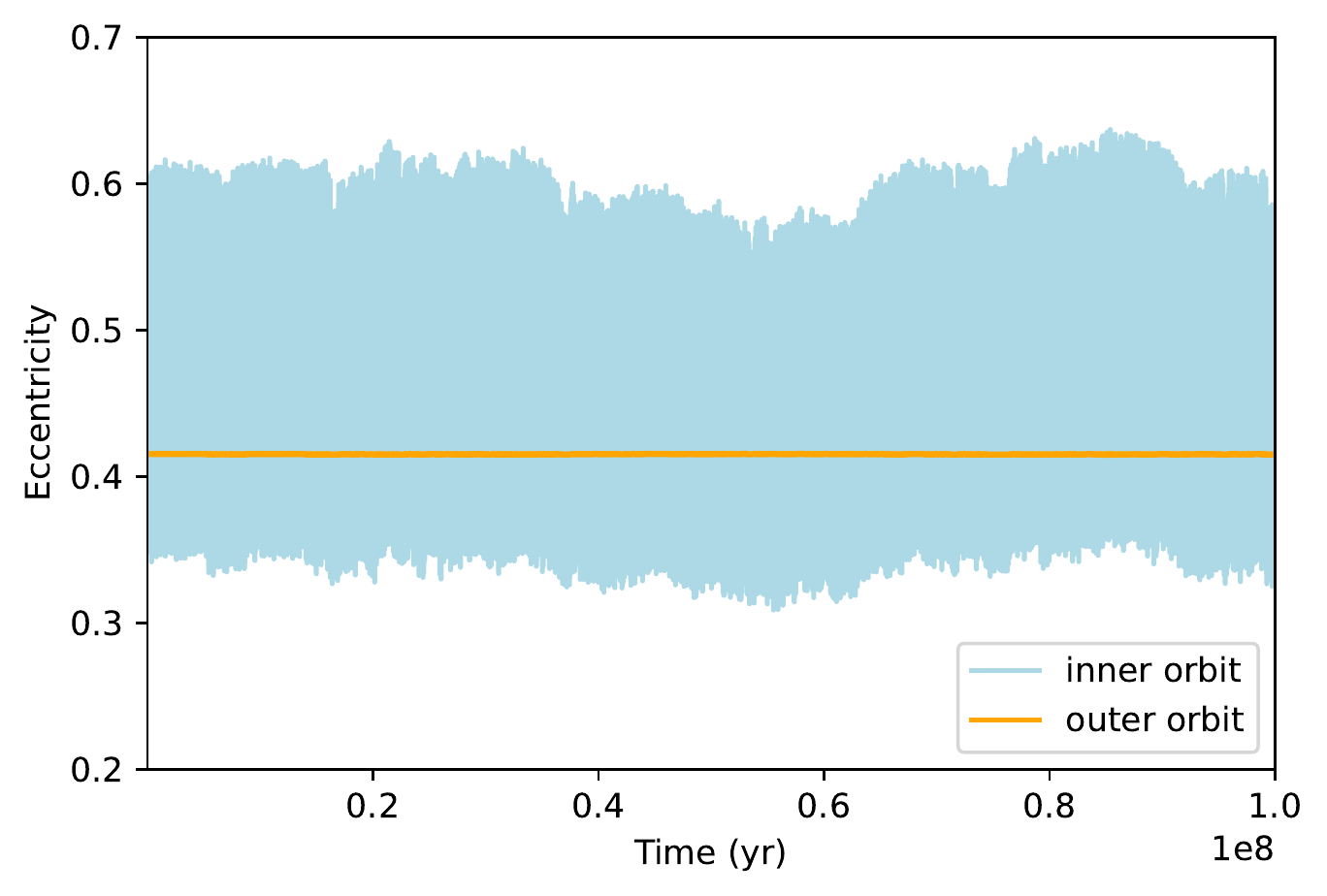}
    \hfil    
    \includegraphics[width=0.3\textwidth]{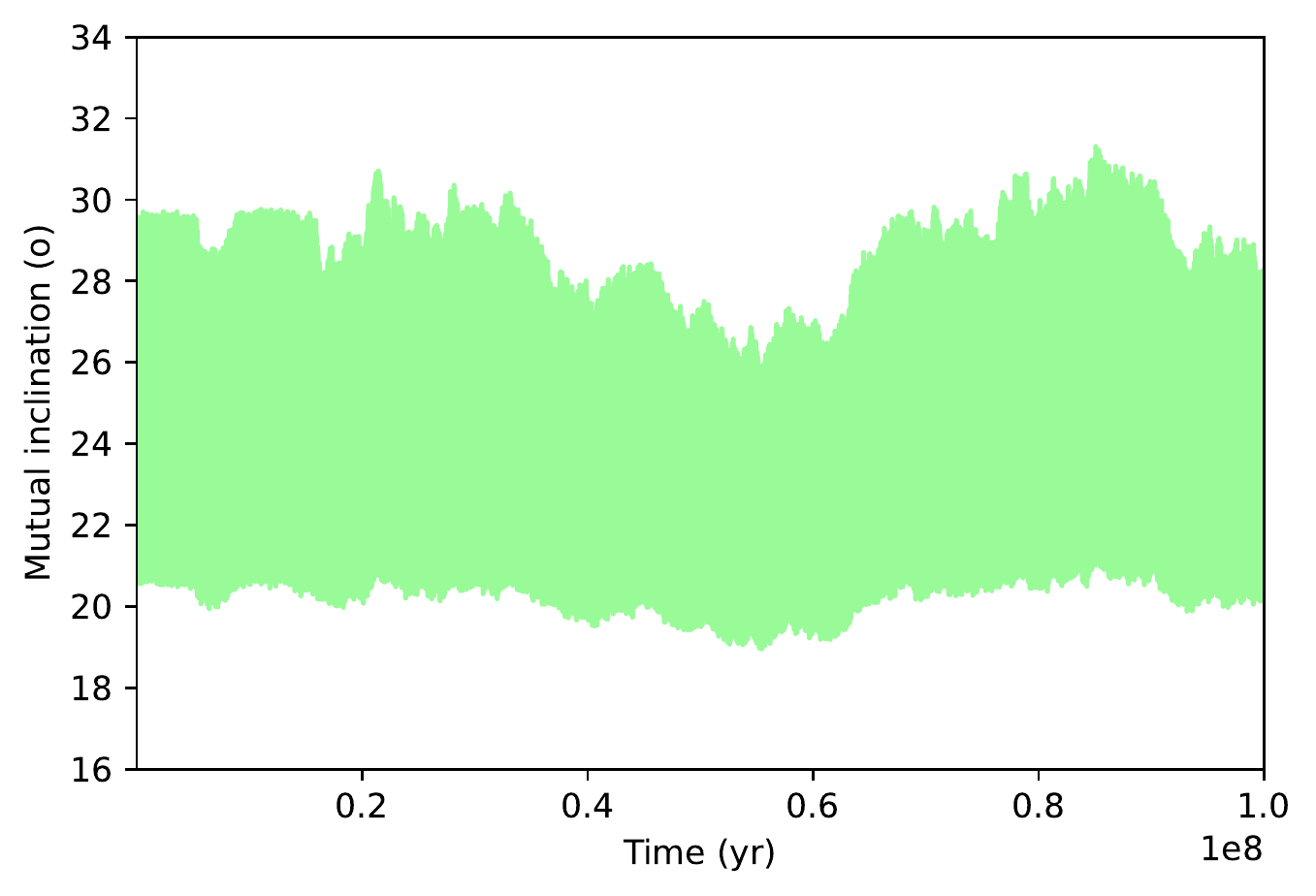}
 }
  \caption{Dynamical evolution of HD\,196885 system as computed with the HJS integrator starting from the orbital solution of Table~\ref{tab:params} over $10^8\,$yr. \textit{Left-Right}: Semi-major axes, eccentricities, and mutual inclination. In the \textit{Left} and \textit{Central} plots, the \textit{orange} curves stand for the inner orbit and the \textit{blue} one for the outer orbit, respectively.} 
    \label{fig:dynsol}
\end{figure*}

\section{Origin and fate\label{sec:stability}}
Our orbital determination clearly suggests that the 3-body system HD\,196885 is presumably non-coplanar. Figure~\ref{fig:params} shows the current favored family of retrograde solutions for which mutual inclination $\Phi$ lies between typical values of 27.6\,deg (first quartile) and 37.8\,deg (third quartile) (note that the MAP value is equal to 24.4\,deg). This non negligible inclination makes it a good candidate for a von Zipele Kozai Lidov-Lidov mechanism (or resonance) \cite{1962AJ.....67..591K} mainly affecting the inner orbit. This dynamical behavior is common in non-planar multiple systems \citep{2006A&A...446..137B,1999MNRAS.304..720K,2000ApJ...535..385F}. It is characterized by coupled large amplitude eccentricity and inclination oscillations mainly affecting the inner orbit. Regarding the relatively large individual masses involved here, this could lead the system to instability, which justifies a dedicated stability study.

To do it, we used the HJS integrator \citep{2003A&A...400.1129B}, a symplectic package specifically designed to study hierarchical stellar systems. The same package was successfully indeed used by \citet{2006A&A...446..137B} to investigate the multiple system GG Tauri. Here we performed an integration over $10^8\,$yr starting from the orbital solution from Table~\ref{tab:params}. The time step was fixed to 0.05\,yr, which is a very conservative assumption, as this represents 1/70 of the smaller orbital period (AaAb orbit), while tests by \citet{2003A&A...400.1129B} show that with only 1/20, the total energy is preserved to $10^{-6}$ relative fraction. But as we fear that the system might be close to instability given the high individual masses of the various bodies involved, this appeared to be a reasonable assumption. Figure~\ref{fig:dynsol} shows the result of this computation, taking the orbital fitting solution from Table~\ref{tab:params} as starting point. The evolution is shown over $10^8\,$yr. This configuration appears stable over the whole integration, as can be seen from the stability of the semi-major axes. Nonetheless the eccentricity and mutual inclination plots suggest von Zipele Kozai Lidov cycles that affect the inner orbit, while the outer orbit is only little affected. This von Zipele Kozai Lidov resonance remains moderate here, due to the rather small mutual inclination. Hence the stability of the triple system is not affected.\\
The formation mechanism of such a system is unclear. We currently do not know if planets in such tight binaries follow similar or different formation processes than the ones orbiting single stars. For instance, \cite{2022MNRAS.511..457C} 
have investigated the role of binary companions triggering fragmentation in self-gravitating discs. They find that binarity triggers disk instability for intermediate-close (100-400\,au) binaries while the presence of a binary companion within 100\,au has a strong negative role against the occurrence of instability. For core accretion, in the specific case of HD\,196885\,b, \cite{2011CeMDA.111...29T} conclude that the perturbed dynamical environment is strongly hostile to planetesimal accretion in the region where the planet is located today (note that this study considered a coplanar case). 
Finally, \cite{2007A&A...467..347M} propose that the present dynamical configuration may not reflect the formation process since the binary orbit may have changed in the past once the planet formed. Disrupted triple system or single stellar encounter could be alternative processes. Consequently, this formation and dynamical picture must be further re-investigated in the light of upcoming astrometric observations and more detailed simulations. First, the solution depicted in Table~\ref{tab:params} may not be unique, and a dedicated follow-up with GRAVITY should enable to rapidly secure the most probable family of solutions. Due to the complexity of the parameter space, other configurations might fit the data as well. Second, other effects could need to be taken into account such as a combination of von Zipele Kozai Lidov cycles and tides \cite[e.g.][]{2007ApJ...669.1298F,2011CeMDA.111..105C,2012A&A...545A..88B}.
    
\section{Conclusion\label{sec:conclusion}}

In this study, we have revisited the global architecture of the extreme planetary system HD\,196885. This system is composed of a tight F8V-M1V binary hosting a S-type circumprimary planet around the A component. Based on the combination of new  measurements in radial velocity, speckle interferometry, high-contrast imaging and high-precision astrometry with
interferometry, we could narrow down the orbital properties of both HD\,196885\,B and HD\,196885\,Ab. The binary companion orbits on an inclined and retrograde plan ($i_{\rm AB}=120.43$\,deg) with a semi-major axis of 19.78\,au, and an eccentricity of 0.415 tightly constrained by almost 40\,yrs and 20\,yrs of radial and astrometric monitoring, respectively. For the radial velocity planet HD\,196885\,Ab, we unambiguously confirm its mass ($M_{\rm Ab}=3.39$\,M$_{\rm Jup}$) and planetary nature, rejecting all families of pole-on solutions corresponding to stellar and brown dwarf masses. The most favored island of orbital solutions found is associated with moderate eccentricity ($e_{\rm AaAb}=0.46$), and moderate-inclination close to $143.04$\,deg. This results points toward a moderate mutual inclination ($\Phi=24.36$\,deg) between the orbital plans of B and Ab. This configuration seems to be dynamically stable over time, although the eccentricity and mutual inclination variations clearly reveal the presence of moderate von Zipele Kozai Lidov cycles that may affect the inner planet. Further observations in the coming years will allow to reject remaining islands of orbital solutions for Ab. Further dynamical simulation should also enable to incorporate additional effects as tides to assess the inner orbit dynamical evolution and stability. 

\begin{acknowledgements} We would like to thank the staﬀ of ESO VLT/I for their support at the telescope at Paranal and La Silla, and the preparation of the observation at Garching. This publication made use of the SIMBAD and VizieR database operated at the CDS, Strasbourg, France.  This work has made use of data from the European Space Agency (ESA) mission Gaia (https://www.cosmos.esa.int/gaia), processed by the Gaia Data Processing and Analysis Consortium (DPAC, https://www.cosmos.esa.int/web/gaia/dpac/consortium). Funding for the DPAC has been provided by national institutions, in particular the institutions participating in the Gaia Multilateral Agreement.
This work is partly based on data products produced at the SPHERE Data Centre hosted at OSUG/IPAG, Grenoble. We thank P. Delorme and E. Lagadec (SPHERE Data Centre) for their efficient help during the data reduction process. RAM and MV acknowledge partial support from FONDECYT/ANID Grant No. 1190038 and to the Chilean National Time Allocation Committee for their continuous support of our SOAR Speckle survey of southern binaries and multiple systems through programs CN2018A-1, CN2019A-2, CN2019B-13, and CN2020A-19, CN2020B-10, and CN2021B-17. CV and MV acknowledge support from the visiting program of the CNRS French-Chilean Laboratory for Astrophysics (IRL-3386). VF acknowledges funding from the National Aeronautics and Space Administration through the Exoplanet Research Program under Grant No. 80NSSC21K0394 (PI: S. Ertel). Based in part on radial velocities obtained with the
SOPHIE spectrograph mounted on the 1.93 m telescope at Observatoire de
Haute-Provence. We warmly thank the OHP staff for their support on the
observations. I.B. received funding from the French Programme National de Planétologie (PNP) of CNRS (INSU). 
A.C. acknowledges support from CFisUC (UIDB/04564/2020 and UIDP/04564/2020), GRAVITY (PTDC/FIS-AST/7002/2020), PHOBOS (POCI-01-0145-FEDER-029932), and ENGAGE~SKA (POCI-01-0145-FEDER-022217), funded by COMPETE 2020 and FCT, Portugal.
JSJ greatfully acknowledges support by FONDECYT grant 1201371 and from the ANID BASAL projects ACE210002 and FB210003. A.G. acknowledges support from the ANID-ALMA fund No. ASTRO20-0059\end{acknowledgements}

%
%

\bibliographystyle{aa.bst}
\bibliography{chauvin-hd196885}
\begin{appendix}

\section{Families of three-bodies fitting solutions using only astrometric datasets from SOAR, NaCo, SPHERE and GRAVITY}

\begin{figure}[ht]
     \centering \includegraphics[width=\columnwidth]{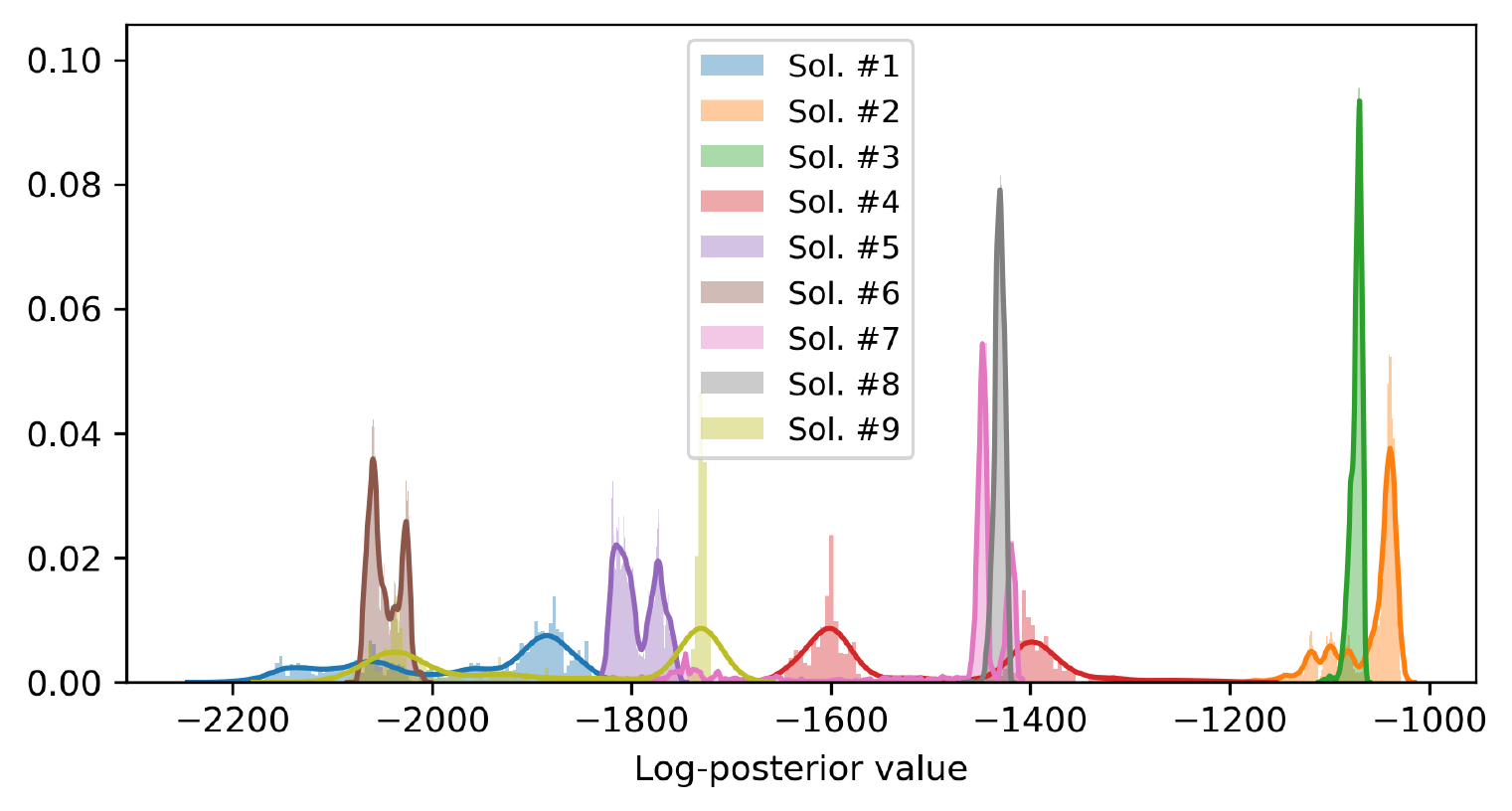}
         \caption{Log-posterior density (up to a constant) distributions of 9 computed inferences considering truncated Gaussian priors on the stellar masses.} 
    \label{fig:logpost_trunc}
\end{figure}

In Figure\,\ref{fig:logpost_trunc} are reported the log posterior values of the different orbital fitting solutions found for the truncated gaussian prior for the mass of the primary. In Table\,\ref{tab:sol_trunc} are summarized the orbital solutions found for these different families.

\begin{table*}[th]
    \label{tab:sol_trunc}
   \centering
    \small
         \caption{Families of solutions for the truncated gaussian prior for the mass of HD\,196885\,A.}
\begin{tabular}{lllllllll}
\hline\hline \noalign{\smallskip}
Sol.  &     $M_{\rm{Aa}}$ &            $M_{\rm{Ab}}$ &               $e_{\rm{AaAb}}$ &               $i_{\rm{AaAb}}$ &      $M_{\rm{B}}$ &                 $e_{\rm{AB}}$ &                 $i_{\rm{AB}}$ &                   $\Phi$ \\
\# & ($\rm{M}_\odot$) & ($\rm{M}_{\rm{Jup}}$) &  & ($^\circ$) & ($\rm{M}_\odot$) &  & ($^\circ$) & ($^\circ$) \\
\hline \noalign{\smallskip}
     1 &  $1.113_{1.111}^{1.113}$ &     $3.357_{3.375}^{8.793}$ &  $0.038_{0.060}^{0.117}$ &     $31.107_{10.883}^{30.922}$ &  $0.526_{0.528}^{0.530}$ &  $0.406_{0.385}^{0.400}$ &  $119.964_{119.073}^{119.727}$ &  $117.138_{115.951}^{119.184}$ \\ \noalign{\smallskip}
     2 &  $1.111_{1.112}^{1.114}$ &     $3.394_{3.130}^{4.096}$ &  $0.444_{0.439}^{0.457}$ &  $143.041_{138.459}^{149.613}$ &  $0.508_{0.506}^{0.508}$ &  $0.417_{0.413}^{0.418}$ &  $120.427_{120.183}^{120.407}$ &     $24.358_{27.596}^{37.747}$ \\ \noalign{\smallskip}
     3 &  $1.112_{1.112}^{1.114}$ &     $4.877_{2.501}^{4.274}$ &  $0.430_{0.428}^{0.443}$ &  $155.006_{125.213}^{151.312}$ &  $0.511_{0.509}^{0.511}$ &  $0.417_{0.413}^{0.419}$ &  $120.680_{120.422}^{120.669}$ &     $58.363_{58.865}^{66.608}$ \\ \noalign{\smallskip}
     4 &  $1.111_{1.111}^{1.112}$ &     $4.647_{5.481}^{9.764}$ &  $0.149_{0.011}^{0.017}$ &  $152.503_{158.225}^{168.088}$ &  $0.523_{0.525}^{0.534}$ &  $0.410_{0.410}^{0.418}$ &  $120.531_{120.489}^{121.003}$ &     $65.833_{65.140}^{69.971}$ \\ \noalign{\smallskip}
     5 &  $1.111_{1.111}^{1.111}$ &     $7.144_{7.125}^{7.202}$ &  $0.402_{0.401}^{0.403}$ &     $14.256_{14.266}^{14.378}$ &  $0.537_{0.534}^{0.536}$ &  $0.382_{0.384}^{0.387}$ &  $119.691_{119.658}^{119.800}$ &  $131.891_{131.863}^{132.162}$ \\ \noalign{\smallskip}
     6 &  $1.111_{1.111}^{1.112}$ &  $12.129_{12.349}^{12.744}$ &  $0.319_{0.326}^{0.342}$ &        $8.872_{8.515}^{8.770}$ &  $0.550_{0.550}^{0.552}$ &  $0.369_{0.369}^{0.371}$ &  $119.032_{119.010}^{119.186}$ &  $125.774_{125.910}^{126.198}$ \\ \noalign{\smallskip}
     7 &  $1.111_{1.111}^{1.112}$ &     $3.497_{2.372}^{3.639}$ &  $0.453_{0.451}^{0.481}$ &   $144.252_{55.234}^{145.394}$ &  $0.538_{0.538}^{0.542}$ &  $0.385_{0.382}^{0.389}$ &  $119.218_{119.145}^{119.470}$ &     $25.696_{54.869}^{96.359}$ \\ \noalign{\smallskip}
     8 &  $1.111_{1.111}^{1.112}$ &    $8.922_{9.901}^{10.630}$ &  $0.459_{0.458}^{0.477}$ &     $13.124_{10.861}^{11.745}$ &  $0.540_{0.539}^{0.541}$ &  $0.401_{0.404}^{0.414}$ &  $119.544_{119.649}^{120.042}$ &  $111.104_{112.750}^{114.008}$ \\ \noalign{\smallskip}
     9 &  $1.111_{1.111}^{1.112}$ &     $3.657_{3.654}^{4.017}$ &  $0.549_{0.439}^{0.569}$ &  $151.370_{151.634}^{155.523}$ &  $0.539_{0.539}^{0.543}$ &  $0.412_{0.398}^{0.412}$ &  $120.230_{119.703}^{120.242}$ &     $88.322_{84.440}^{88.129}$ \\ \noalign{\smallskip}
\hline
\end{tabular}
\end{table*}

\end{appendix}

\end{document}

%% file: summary_table.tex
\begin{tabular}{lll}
\hline\hline \noalign{\smallskip}
             Parameters &                                Ab &                                 B \\
\hline \noalign{\smallskip}
               $a$ (au) &           $2.383_{2.379}^{2.385}$ &        $19.778_{19.759}^{19.886}$ \\ \noalign{\smallskip}
               $P$ (yr) &           $3.485_{3.469}^{3.486}$ &        $69.045_{68.934}^{69.578}$ \\ \noalign{\smallskip}
                    $e$ &           $0.444_{0.439}^{0.457}$ &           $0.417_{0.413}^{0.418}$ \\ \noalign{\smallskip}
         $i$ ($^\circ$) &     $143.041_{138.459}^{149.613}$ &     $120.427_{120.183}^{120.407}$ \\ \noalign{\smallskip}
    $\Omega$ ($^\circ$) &      $91.570_{103.696}^{116.435}$ &        $79.150_{78.974}^{79.138}$ \\ \noalign{\smallskip}
    $\omega$ ($^\circ$) &     $150.166_{153.034}^{159.668}$ &     $231.464_{231.503}^{232.036}$ \\ \noalign{\smallskip}
          $t_P$ (yr AD) &  $1982.517_{1982.525}^{1982.732}$ &  $1982.886_{1982.832}^{1982.985}$ \\ \noalign{\smallskip}
 $M_{\rm{Aa}}$ ($\rm{M}_{\odot}$) &           $1.111_{1.112}^{1.114}$ &                                   \\ \noalign{\smallskip}
  $M_{\rm{Ab}}$ ($\rm{M}_{\rm{Jup}}$)  &           $3.394_{3.130}^{4.096}$ &                                   \\ \noalign{\smallskip}
 $M_{\rm{B}}$ ($\rm{M}_{\odot}$)  &                                   &           $0.508_{0.506}^{0.508}$ \\ \noalign{\smallskip}
      $\Phi$ ($^\circ$) &           \multicolumn{2}{c}{$24.358_{27.596}^{37.747}$}              \\ \noalign{\smallskip}
\hline
\end{tabular}

%% file: chauvin-hd196885.bbl
\begin{thebibliography}{88}
\expandafter\ifx\csname natexlab\endcsname\relax\def\natexlab#1{#1}\fi

\bibitem[{{Abushattal} {et~al.}(2020){Abushattal}, {Docobo}, \&
  {Campo}}]{Abuset20}
{Abushattal}, A.~A., {Docobo}, J.~A., \& {Campo}, P.~P. 2020, \aj, 159, 28

\bibitem[{{Baranec} {et~al.}(2016){Baranec}, {Ziegler}, {Law}, {Morton},
  {Riddle}, {Atkinson}, {Schonhut}, \& {Crepp}}]{2016AJ....152...18B}
{Baranec}, C., {Ziegler}, C., {Law}, N.~M., {et~al.} 2016, \aj, 152, 18

\bibitem[{{Bazs{\'o}} {et~al.}(2017){Bazs{\'o}}, {Pilat-Lohinger}, {Eggl},
  {Funk}, {Bancelin}, \& {Rau}}]{2017MNRAS.466.1555B}
{Bazs{\'o}}, {\'A}., {Pilat-Lohinger}, E., {Eggl}, S., {et~al.} 2017, \mnras,
  466, 1555

\bibitem[{{Benedict} {et~al.}(2018){Benedict}, {Harrison}, {Endl}, \&
  {Torres}}]{2018RNAAS...2....7B}
{Benedict}, G.~F., {Harrison}, T.~E., {Endl}, M., \& {Torres}, G. 2018,
  Research Notes of the American Astronomical Society, 2, 7

\bibitem[{{Beust}(2003)}]{2003A&A...400.1129B}
{Beust}, H. 2003, \aap, 400, 1129

\bibitem[{{Beust} {et~al.}(2012){Beust}, {Bonfils}, {Montagnier}, {Delfosse},
  \& {Forveille}}]{2012A&A...545A..88B}
{Beust}, H., {Bonfils}, X., {Montagnier}, G., {Delfosse}, X., \& {Forveille},
  T. 2012, \aap, 545, A88

\bibitem[{{Beust} \& {Dutrey}(2006)}]{2006A&A...446..137B}
{Beust}, H. \& {Dutrey}, A. 2006, \aap, 446, 137

\bibitem[{{Beuzit} {et~al.}(2019){Beuzit}, {Vigan}, {Mouillet}, {Dohlen},
  {Gratton}, {Boccaletti}, {Sauvage}, {Schmid}, {Langlois}, \&
  {Petit}}]{Beuzit2019_sphere}
{Beuzit}, J.~L., {Vigan}, A., {Mouillet}, D., {et~al.} 2019, arXiv e-prints,
  arXiv:1902.04080

\bibitem[{{Bonavita} \& {Desidera}(2007)}]{2007A&A...468..721B}
{Bonavita}, M. \& {Desidera}, S. 2007, \aap, 468, 721

\bibitem[{{Bonavita} \& {Desidera}(2020)}]{2020Galax...8...16B}
{Bonavita}, M. \& {Desidera}, S. 2020, Galaxies, 8, 16

\bibitem[{{Bouchy} {et~al.}(2013){Bouchy}, {D{\'\i}az}, {H{\'e}brard},
  {Arnold}, {Boisse}, {Delfosse}, {Perruchot}, \&
  {Santerne}}]{2013A&A...549A..49B}
{Bouchy}, F., {D{\'\i}az}, R.~F., {H{\'e}brard}, G., {et~al.} 2013, \aap, 549,
  A49

\bibitem[{{Bouchy} {et~al.}(2009){Bouchy}, {H{\'e}brard}, {Udry}, {Delfosse},
  {Boisse}, {Desort}, {Bonfils}, {Eggenberger}, {Ehrenreich}, {Forveille},
  {Lagrange}, {Le Coroller}, {Lovis}, {Moutou}, {Pepe}, {Perrier}, {Pont},
  {Queloz}, {Santos}, {S{\'e}gransan}, \& {Vidal-Madjar}}]{2009A&A...505..853B}
{Bouchy}, F., {H{\'e}brard}, G., {Udry}, S., {et~al.} 2009, \aap, 505, 853

\bibitem[{{Cadman} {et~al.}(2022){Cadman}, {Hall}, {Fontanive}, \&
  {Rice}}]{2022MNRAS.511..457C}
{Cadman}, J., {Hall}, C., {Fontanive}, C., \& {Rice}, K. 2022, \mnras, 511, 457

\bibitem[{Carpenter {et~al.}(2017)Carpenter, Gelman, Hoffman, Lee, Goodrich,
  Betancourt, Brubaker, Guo, Li, \& Riddell}]{carpenter2017stan}
Carpenter, B., Gelman, A., Hoffman, M.~D., {et~al.} 2017, Journal of
  statistical software, 76, 1

\bibitem[{{Chauvin} {et~al.}(2011){Chauvin}, {Beust}, {Lagrange}, \&
  {Eggenberger}}]{chauvin2011}
{Chauvin}, G., {Beust}, H., {Lagrange}, A.~M., \& {Eggenberger}, A. 2011, \aap,
  528, A8

\bibitem[{{Chauvin} {et~al.}(2017){Chauvin}, {Desidera}, {Lagrange}, {Vigan},
  {Feldt}, {Gratton}, {Langlois}, {Cheetham}, {Bonnefoy}, \&
  {Meyer}}]{Chauvin2017_shine}
{Chauvin}, G., {Desidera}, S., {Lagrange}, A.~M., {et~al.} 2017, in SF2A-2017:
  Proceedings of the Annual meeting of the French Society of Astronomy and
  Astrophysics, Di

\bibitem[{{Chauvin} {et~al.}(2006){Chauvin}, {Lagrange}, {Udry}, {Fusco},
  {Galland}, {Naef}, {Beuzit}, \& {Mayor}}]{2006A&A...456.1165C}
{Chauvin}, G., {Lagrange}, A.~M., {Udry}, S., {et~al.} 2006, \aap, 456, 1165

\bibitem[{{Chauvin} {et~al.}(2007){Chauvin}, {Lagrange}, {Udry}, \&
  {Mayor}}]{2007A&A...475..723C}
{Chauvin}, G., {Lagrange}, A.~M., {Udry}, S., \& {Mayor}, M. 2007, \aap, 475,
  723

\bibitem[{{Claudi} {et~al.}(2008){Claudi}, {Turatto}, {Gratton}, {Antichi},
  {Bonavita}, {Bruno}, {Cascone}, {De Caprio}, {Desidera}, {Giro}, {Mesa},
  {Scuderi}, {Dohlen}, {Beuzit}, \& {Puget}}]{Claudi2008}
{Claudi}, R.~U., {Turatto}, M., {Gratton}, R.~G., {et~al.} 2008, in Society of
  Photo-Optical Instrumentation Engineers (SPIE) Conference Series, Vol. 7014,
  Society of Photo-Optical Instrumentation Engineers (SPIE) Conference Series

\bibitem[{{Correia} {et~al.}(2011){Correia}, {Laskar}, {Farago}, \&
  {Bou{\'e}}}]{2011CeMDA.111..105C}
{Correia}, A. C.~M., {Laskar}, J., {Farago}, F., \& {Bou{\'e}}, G. 2011,
  Celestial Mechanics and Dynamical Astronomy, 111, 105

\bibitem[{{Correia} {et~al.}(2008){Correia}, {Udry}, {Mayor}, {Eggenberger},
  {Naef}, {Beuzit}, {Perrier}, {Queloz}, {Sivan}, {Pepe}, {Santos}, \&
  {S{\'e}gransan}}]{2008A&A...479..271C}
{Correia}, A.~C.~M., {Udry}, S., {Mayor}, M., {et~al.} 2008, \aap, 479, 271

\bibitem[{{Desidera} \& {Barbieri}(2007)}]{2007A&A...462..345D}
{Desidera}, S. \& {Barbieri}, M. 2007, \aap, 462, 345

\bibitem[{{Desidera} {et~al.}(2021){Desidera}, {Chauvin}, {Bonavita},
  {Messina}, {LeCoroller}, {Schmidt}, {Gratton}, {Lazzoni}, {Meyer},
  {Schlieder}, {Cheetham}, {Hagelberg}, {Bonnefoy}, {Feldt}, {Lagrange},
  {Langlois}, {Vigan}, {Tan}, {Hambsch}, {Millward}, {Alcal{\'a}}, {Benatti},
  {Brandner}, {Carson}, {Covino}, {Delorme}, {D'Orazi}, {Janson}, {Rigliaco},
  {Beuzit}, {Biller}, {Boccaletti}, {Dominik}, {Cantalloube}, {Fontanive},
  {Galicher}, {Henning}, {Lagadec}, {Ligi}, {Maire}, {Menard}, {Mesa},
  {M{\"u}ller}, {Samland}, {Schmid}, {Sissa}, {Turatto}, {Udry}, {Zurlo},
  {Asensio-Torres}, {Kopytova}, {Rickman}, {Abe}, {Antichi}, {Baruffolo},
  {Baudoz}, {Baudrand}, {Blanchard}, {Bazzon}, {Buey}, {Carbillet}, {Carle},
  {Charton}, {Cascone}, {Claudi}, {Costille}, {Deboulb{\'e}}, {De Caprio},
  {Dohlen}, {Fantinel}, {Feautrier}, {Fusco}, {Gigan}, {Giro}, {Gisler},
  {Gluck}, {Hubin}, {Hugot}, {Jaquet}, {Kasper}, {Madec}, {Magnard},
  {Martinez}, {Maurel}, {Le Mignant}, {M{\"o}ller-Nilsson}, {Llored}, {Moulin},
  {Orign{\'e}}, {Pavlov}, {Perret}, {Petit}, {Pragt}, {Puget}, {Rabou},
  {Ramos}, {Rigal}, {Rochat}, {Roelfsema}, {Rousset}, {Roux}, {Salasnich},
  {Sauvage}, {Sevin}, {Soenke}, {Stadler}, {Suarez}, {Weber}, \&
  {Wildi}}]{Desidera2021_shine_paperI_sample_definition}
{Desidera}, S., {Chauvin}, G., {Bonavita}, M., {et~al.} 2021, \aap, 651, A70

\bibitem[{{Desidera} {et~al.}(2010){Desidera}, {Gratton}, {Endl}, {Fiorenzano},
  {Barbieri}, {Claudi}, {Cosentino}, {Scuderi}, \&
  {Bonavita}}]{2010ASSL..366..105D}
{Desidera}, S., {Gratton}, R., {Endl}, M., {et~al.} 2010, in Astrophysics and
  Space Science Library, Vol. 366, Planets in Binary Star Systems, ed.
  N.~{Haghighipour}, 105

\bibitem[{{Dohlen} {et~al.}(2008){Dohlen}, {Langlois}, {Saisse}, {Hill},
  {Origne}, {Jacquet}, {Fabron}, {Blanc}, {Llored}, {Carle}, {Moutou}, {Vigan},
  {Boccaletti}, {Carbillet}, {Mouillet}, \& {Beuzit}}]{Dohlen2008}
{Dohlen}, K., {Langlois}, M., {Saisse}, M., {et~al.} 2008, in \procspie, Vol.
  7014, Ground-based and Airborne Instrumentation for Astronomy II, 70143L

\bibitem[{{Duch{\^e}ne}(2010)}]{2010ApJ...709L.114D}
{Duch{\^e}ne}, G. 2010, \apjl, 709, L114

\bibitem[{{Dvorak}(1982)}]{1982OAWMN.191..423D}
{Dvorak}, R. 1982, Oesterreichische Akademie Wissenschaften Mathematisch
  naturwissenschaftliche Klasse Sitzungsberichte Abteilung, 191, 423

\bibitem[{{Eggenberger} \& {Udry}(2007)}]{2007arXiv0705.3173E}
{Eggenberger}, A. \& {Udry}, S. 2007, arXiv e-prints, arXiv:0705.3173

\bibitem[{{Eggenberger} {et~al.}(2008){Eggenberger}, {Udry}, {Chauvin},
  {Beuzit}, {Lagrange}, \& {Mayor}}]{2008ASPC..398..179E}
{Eggenberger}, A., {Udry}, S., {Chauvin}, G., {et~al.} 2008, in Astronomical
  Society of the Pacific Conference Series, Vol. 398, Extreme Solar Systems,
  ed. D.~{Fischer}, F.~A. {Rasio}, S.~E. {Thorsett}, \& A.~{Wolszczan}, 179

\bibitem[{{Eggenberger} {et~al.}(2007){Eggenberger}, {Udry}, {Chauvin},
  {Beuzit}, {Lagrange}, {S{\'e}gransan}, \& {Mayor}}]{2007A&A...474..273E}
{Eggenberger}, A., {Udry}, S., {Chauvin}, G., {et~al.} 2007, \aap, 474, 273

\bibitem[{{Eggenberger} {et~al.}(2004){Eggenberger}, {Udry}, {Mayor}, {Beuzit},
  {Lagrange}, \& {Chauvin}}]{2004ASPC..321...93E}
{Eggenberger}, A., {Udry}, S., {Mayor}, M., {et~al.} 2004, in Astronomical
  Society of the Pacific Conference Series, Vol. 321, Extrasolar Planets: Today
  and Tomorrow, ed. J.~{Beaulieu}, A.~{Lecavelier Des Etangs}, \& C.~{Terquem},
  93

\bibitem[{{Fabrycky} \& {Tremaine}(2007)}]{2007ApJ...669.1298F}
{Fabrycky}, D. \& {Tremaine}, S. 2007, \apj, 669, 1298

\bibitem[{{Fischer} {et~al.}(2009){Fischer}, {Driscoll}, {Isaacson}, {Giguere},
  {Marcy}, {Valenti}, {Wright}, {Henry}, {Johnson}, {Howard}, {Peek}, \&
  {McCarthy}}]{2009ApJ...703.1545F}
{Fischer}, D., {Driscoll}, P., {Isaacson}, H., {et~al.} 2009, \apj, 703, 1545

\bibitem[{{Fontanive} \& {Bardalez Gagliuffi}(2021)}]{2021FrASS...8...16F}
{Fontanive}, C. \& {Bardalez Gagliuffi}, D. 2021, Frontiers in Astronomy and
  Space Sciences, 8, 16

\bibitem[{Ford(2005)}]{ford2005quantifying}
Ford, E.~B. 2005, The Astronomical Journal, 129, 1706

\bibitem[{{Ford} {et~al.}(2000){Ford}, {Kozinsky}, \&
  {Rasio}}]{2000ApJ...535..385F}
{Ford}, E.~B., {Kozinsky}, B., \& {Rasio}, F.~A. 2000, \apj, 535, 385

\bibitem[{{Fragner} {et~al.}(2011){Fragner}, {Nelson}, \&
  {Kley}}]{2011A&A...528A..40F}
{Fragner}, M.~M., {Nelson}, R.~P., \& {Kley}, W. 2011, \aap, 528, A40

\bibitem[{{Gaia Collaboration} {et~al.}(2018){Gaia Collaboration}, {Brown},
  {Vallenari}, {Prusti}, {de Bruijne}, {Babusiaux}, {Bailer-Jones}, {Biermann},
  {Evans}, {Eyer}, {Jansen}, {Jordi}, {Klioner}, {Lammers}, {Lindegren},
  {Luri}, {Mignard}, {Panem}, {Pourbaix}, {Randich}, {Sartoretti}, {Siddiqui},
  {Soubiran}, {van Leeuwen}, {Walton}, {Arenou}, {Bastian}, {Cropper},
  {Drimmel}, {Katz}, {Lattanzi}, {Bakker}, {Cacciari}, {Casta{\~n}eda},
  {Chaoul}, {Cheek}, {De Angeli}, {Fabricius}, {Guerra}, {Holl}, {Masana},
  {Messineo}, {Mowlavi}, {Nienartowicz}, {Panuzzo}, {Portell}, {Riello},
  {Seabroke}, {Tanga}, {Th{\'e}venin}, {Gracia-Abril}, {Comoretto},
  {Garcia-Reinaldos}, {Teyssier}, {Altmann}, {Andrae}, {Audard},
  {Bellas-Velidis}, {Benson}, {Berthier}, {Blomme}, {Burgess}, {Busso},
  {Carry}, {Cellino}, {Clementini}, {Clotet}, {Creevey}, {Davidson}, {De
  Ridder}, {Delchambre}, {Dell'Oro}, {Ducourant},
  {Fern{\'a}ndez-Hern{\'a}ndez}, {Fouesneau}, {Fr{\'e}mat}, {Galluccio},
  {Garc{\'\i}a-Torres}, {Gonz{\'a}lez-N{\'u}{\~n}ez}, {Gonz{\'a}lez-Vidal},
  {Gosset}, {Guy}, {Halbwachs}, {Hambly}, {Harrison}, {Hern{\'a}ndez},
  {Hestroffer}, {Hodgkin}, {Hutton}, {Jasniewicz}, {Jean-Antoine-Piccolo},
  {Jordan}, {Korn}, {Krone-Martins}, {Lanzafame}, {Lebzelter}, {L{\"o}ffler},
  {Manteiga}, {Marrese}, {Mart{\'\i}n-Fleitas}, {Moitinho}, {Mora}, {Muinonen},
  {Osinde}, {Pancino}, {Pauwels}, {Petit}, {Recio-Blanco}, {Richards},
  {Rimoldini}, {Robin}, {Sarro}, {Siopis}, {Smith}, {Sozzetti}, {S{\"u}veges},
  {Torra}, {van Reeven}, {Abbas}, {Abreu Aramburu}, {Accart}, {Aerts},
  {Altavilla}, {{\'A}lvarez}, {Alvarez}, {Alves}, {Anderson}, {Andrei},
  {Anglada Varela}, {Antiche}, {Antoja}, {Arcay}, {Astraatmadja}, {Bach},
  {Baker}, {Balaguer-N{\'u}{\~n}ez}, {Balm}, {Barache}, {Barata}, {Barbato},
  {Barblan}, {Barklem}, {Barrado}, {Barros}, {Barstow}, {Bartholom{\'e}
  Mu{\~n}oz}, {Bassilana}, {Becciani}, {Bellazzini}, {Berihuete}, {Bertone},
  {Bianchi}, {Bienaym{\'e}}, {Blanco-Cuaresma}, {Boch}, {Boeche}, {Bombrun},
  {Borrachero}, {Bossini}, {Bouquillon}, {Bourda}, {Bragaglia}, {Bramante},
  {Breddels}, {Bressan}, {Brouillet}, {Br{\"u}semeister}, {Brugaletta},
  {Bucciarelli}, {Burlacu}, {Busonero}, {Butkevich}, {Buzzi}, {Caffau},
  {Cancelliere}, {Cannizzaro}, {Cantat-Gaudin}, {Carballo}, {Carlucci},
  {Carrasco}, {Casamiquela}, {Castellani}, {Castro-Ginard}, {Charlot},
  {Chemin}, {Chiavassa}, {Cocozza}, {Costigan}, {Cowell}, {Crifo}, {Crosta},
  {Crowley}, {Cuypers}, {Dafonte}, {Damerdji}, {Dapergolas}, {David}, {David},
  {de Laverny}, {De Luise}, {De March}, {de Martino}, {de Souza}, {de Torres},
  {Debosscher}, {del Pozo}, {Delbo}, {Delgado}, {Delgado}, {Di Matteo},
  {Diakite}, {Diener}, {Distefano}, {Dolding}, {Drazinos}, {Dur{\'a}n},
  {Edvardsson}, {Enke}, {Eriksson}, {Esquej}, {Eynard Bontemps}, {Fabre},
  {Fabrizio}, {Faigler}, {Falc{\~a}o}, {Farr{\`a}s Casas}, {Federici},
  {Fedorets}, {Fernique}, {Figueras}, {Filippi}, {Findeisen}, {Fonti},
  {Fraile}, {Fraser}, {Fr{\'e}zouls}, {Gai}, {Galleti}, {Garabato},
  {Garc{\'\i}a-Sedano}, {Garofalo}, {Garralda}, {Gavel}, {Gavras}, {Gerssen},
  {Geyer}, {Giacobbe}, {Gilmore}, {Girona}, {Giuffrida}, {Glass}, {Gomes},
  {Granvik}, {Gueguen}, {Guerrier}, {Guiraud}, {Guti{\'e}rrez-S{\'a}nchez},
  {Haigron}, {Hatzidimitriou}, {Hauser}, {Haywood}, {Heiter}, {Helmi}, {Heu},
  {Hilger}, {Hobbs}, {Hofmann}, {Holland}, {Huckle}, {Hypki}, {Icardi},
  {Jan{\ss}en}, {Jevardat de Fombelle}, {Jonker}, {Juh{\'a}sz}, {Julbe},
  {Karampelas}, {Kewley}, {Klar}, {Kochoska}, {Kohley}, {Kolenberg},
  {Kontizas}, {Kontizas}, {Koposov}, {Kordopatis}, {Kostrzewa-Rutkowska},
  {Koubsky}, {Lambert}, {Lanza}, {Lasne}, {Lavigne}, {Le Fustec}, {Le
  Poncin-Lafitte}, {Lebreton}, {Leccia}, {Leclerc}, {Lecoeur-Taibi},
  {Lenhardt}, {Leroux}, {Liao}, {Licata}, {Lindstr{\o}m}, {Lister}, {Livanou},
  {Lobel}, {L{\'o}pez}, {Managau}, {Mann}, {Mantelet}, {Marchal}, {Marchant},
  {Marconi}, {Marinoni}, {Marschalk{\'o}}, {Marshall}, {Martino}, {Marton},
  {Mary}, {Massari}, {Matijevi{\v{c}}}, {Mazeh}, {McMillan}, {Messina},
  {Michalik}, {Millar}, {Molina}, {Molinaro}, {Moln{\'a}r}, {Montegriffo},
  {Mor}, {Morbidelli}, {Morel}, {Morris}, {Mulone}, {Muraveva}, {Musella},
  {Nelemans}, {Nicastro}, {Noval}, {O'Mullane}, {Ord{\'e}novic},
  {Ord{\'o}{\~n}ez-Blanco}, {Osborne}, {Pagani}, {Pagano}, {Pailler},
  {Palacin}, {Palaversa}, {Panahi}, {Pawlak}, {Piersimoni}, {Pineau}, {Plachy},
  {Plum}, {Poggio}, {Poujoulet}, {Pr{\v{s}}a}, {Pulone}, {Racero}, {Ragaini},
  {Rambaux}, {Ramos-Lerate}, {Regibo}, {Reyl{\'e}}, {Riclet}, {Ripepi}, {Riva},
  {Rivard}, {Rixon}, {Roegiers}, {Roelens}, {Romero-G{\'o}mez}, {Rowell},
  {Royer}, {Ruiz-Dern}, {Sadowski}, {Sagrist{\`a} Sell{\'e}s}, {Sahlmann},
  {Salgado}, {Salguero}, {Sanna}, {Santana-Ros}, {Sarasso}, {Savietto},
  {Schultheis}, {Sciacca}, {Segol}, {Segovia}, {S{\'e}gransan}, {Shih},
  {Siltala}, {Silva}, {Smart}, {Smith}, {Solano}, {Solitro}, {Sordo}, {Soria
  Nieto}, {Souchay}, {Spagna}, {Spoto}, {Stampa}, {Steele},
  {Steidelm{\"u}ller}, {Stephenson}, {Stoev}, {Suess}, {Surdej}, {Szabados},
  {Szegedi-Elek}, {Tapiador}, {Taris}, {Tauran}, {Taylor}, {Teixeira},
  {Terrett}, {Teyssandier}, {Thuillot}, {Titarenko}, {Torra Clotet}, {Turon},
  {Ulla}, {Utrilla}, {Uzzi}, {Vaillant}, {Valentini}, {Valette}, {van Elteren},
  {Van Hemelryck}, {van Leeuwen}, {Vaschetto}, {Vecchiato}, {Veljanoski},
  {Viala}, {Vicente}, {Vogt}, {von Essen}, {Voss}, {Votruba}, {Voutsinas},
  {Walmsley}, {Weiler}, {Wertz}, {Wevers}, {Wyrzykowski}, {Yoldas},
  {{\v{Z}}erjal}, {Ziaeepour}, {Zorec}, {Zschocke}, {Zucker}, {Zurbach}, \&
  {Zwitter}}]{2018A&A...616A...1G}
{Gaia Collaboration}, {Brown}, A.~G.~A., {Vallenari}, A., {et~al.} 2018, \aap,
  616, A1

\bibitem[{{Gaia Collaboration} {et~al.}(2021){Gaia Collaboration}, {Brown},
  {Vallenari}, {Prusti}, {de Bruijne}, {Babusiaux}, {Biermann}, {Creevey},
  {Evans}, {Eyer}, {Hutton}, {Jansen}, {Jordi}, {Klioner}, {Lammers},
  {Lindegren}, {Luri}, {Mignard}, {Panem}, {Pourbaix}, {Randich}, {Sartoretti},
  {Soubiran}, {Walton}, {Arenou}, {Bailer-Jones}, {Bastian}, {Cropper},
  {Drimmel}, {Katz}, {Lattanzi}, {van Leeuwen}, {Bakker}, {Cacciari},
  {Casta{\~n}eda}, {De Angeli}, {Ducourant}, {Fabricius}, {Fouesneau},
  {Fr{\'e}mat}, {Guerra}, {Guerrier}, {Guiraud}, {Jean-Antoine Piccolo},
  {Masana}, {Messineo}, {Mowlavi}, {Nicolas}, {Nienartowicz}, {Pailler},
  {Panuzzo}, {Riclet}, {Roux}, {Seabroke}, {Sordo}, {Tanga}, {Th{\'e}venin},
  {Gracia-Abril}, {Portell}, {Teyssier}, {Altmann}, {Andrae}, {Bellas-Velidis},
  {Benson}, {Berthier}, {Blomme}, {Brugaletta}, {Burgess}, {Busso}, {Carry},
  {Cellino}, {Cheek}, {Clementini}, {Damerdji}, {Davidson}, {Delchambre},
  {Dell'Oro}, {Fern{\'a}ndez-Hern{\'a}ndez}, {Galluccio}, {Garc{\'\i}a-Lario},
  {Garcia-Reinaldos}, {Gonz{\'a}lez-N{\'u}{\~n}ez}, {Gosset}, {Haigron},
  {Halbwachs}, {Hambly}, {Harrison}, {Hatzidimitriou}, {Heiter},
  {Hern{\'a}ndez}, {Hestroffer}, {Hodgkin}, {Holl}, {Jan{\ss}en}, {Jevardat de
  Fombelle}, {Jordan}, {Krone-Martins}, {Lanzafame}, {L{\"o}ffler}, {Lorca},
  {Manteiga}, {Marchal}, {Marrese}, {Moitinho}, {Mora}, {Muinonen}, {Osborne},
  {Pancino}, {Pauwels}, {Petit}, {Recio-Blanco}, {Richards}, {Riello},
  {Rimoldini}, {Robin}, {Roegiers}, {Rybizki}, {Sarro}, {Siopis}, {Smith},
  {Sozzetti}, {Ulla}, {Utrilla}, {van Leeuwen}, {van Reeven}, {Abbas}, {Abreu
  Aramburu}, {Accart}, {Aerts}, {Aguado}, {Ajaj}, {Altavilla}, {{\'A}lvarez},
  {{\'A}lvarez Cid-Fuentes}, {Alves}, {Anderson}, {Anglada Varela}, {Antoja},
  {Audard}, {Baines}, {Baker}, {Balaguer-N{\'u}{\~n}ez}, {Balbinot}, {Balog},
  {Barache}, {Barbato}, {Barros}, {Barstow}, {Bartolom{\'e}}, {Bassilana},
  {Bauchet}, {Baudesson-Stella}, {Becciani}, {Bellazzini}, {Bernet}, {Bertone},
  {Bianchi}, {Blanco-Cuaresma}, {Boch}, {Bombrun}, {Bossini}, {Bouquillon},
  {Bragaglia}, {Bramante}, {Breedt}, {Bressan}, {Brouillet}, {Bucciarelli},
  {Burlacu}, {Busonero}, {Butkevich}, {Buzzi}, {Caffau}, {Cancelliere},
  {C{\'a}novas}, {Cantat-Gaudin}, {Carballo}, {Carlucci}, {Carnerero},
  {Carrasco}, {Casamiquela}, {Castellani}, {Castro-Ginard}, {Castro Sampol},
  {Chaoul}, {Charlot}, {Chemin}, {Chiavassa}, {Cioni}, {Comoretto}, {Cooper},
  {Cornez}, {Cowell}, {Crifo}, {Crosta}, {Crowley}, {Dafonte}, {Dapergolas},
  {David}, {David}, {de Laverny}, {De Luise}, {De March}, {De Ridder}, {de
  Souza}, {de Teodoro}, {de Torres}, {del Peloso}, {del Pozo}, {Delbo},
  {Delgado}, {Delgado}, {Delisle}, {Di Matteo}, {Diakite}, {Diener},
  {Distefano}, {Dolding}, {Eappachen}, {Edvardsson}, {Enke}, {Esquej}, {Fabre},
  {Fabrizio}, {Faigler}, {Fedorets}, {Fernique}, {Fienga}, {Figueras},
  {Fouron}, {Fragkoudi}, {Fraile}, {Franke}, {Gai}, {Garabato},
  {Garcia-Gutierrez}, {Garc{\'\i}a-Torres}, {Garofalo}, {Gavras}, {Gerlach},
  {Geyer}, {Giacobbe}, {Gilmore}, {Girona}, {Giuffrida}, {Gomel}, {Gomez},
  {Gonzalez-Santamaria}, {Gonz{\'a}lez-Vidal}, {Granvik},
  {Guti{\'e}rrez-S{\'a}nchez}, {Guy}, {Hauser}, {Haywood}, {Helmi}, {Hidalgo},
  {Hilger}, {H{\l}adczuk}, {Hobbs}, {Holland}, {Huckle}, {Jasniewicz},
  {Jonker}, {Juaristi Campillo}, {Julbe}, {Karbevska}, {Kervella}, {Khanna},
  {Kochoska}, {Kontizas}, {Kordopatis}, {Korn}, {Kostrzewa-Rutkowska},
  {Kruszy{\'n}ska}, {Lambert}, {Lanza}, {Lasne}, {Le Campion}, {Le Fustec},
  {Lebreton}, {Lebzelter}, {Leccia}, {Leclerc}, {Lecoeur-Taibi}, {Liao},
  {Licata}, {Lindstr{\o}m}, {Lister}, {Livanou}, {Lobel}, {Madrero Pardo},
  {Managau}, {Mann}, {Marchant}, {Marconi}, {Marcos Santos}, {Marinoni},
  {Marocco}, {Marshall}, {Martin Polo}, {Mart{\'\i}n-Fleitas}, {Masip},
  {Massari}, {Mastrobuono-Battisti}, {Mazeh}, {McMillan}, {Messina},
  {Michalik}, {Millar}, {Mints}, {Molina}, {Molinaro}, {Moln{\'a}r},
  {Montegriffo}, {Mor}, {Morbidelli}, {Morel}, {Morris}, {Mulone}, {Munoz},
  {Muraveva}, {Murphy}, {Musella}, {Noval}, {Ord{\'e}novic}, {Orr{\`u}},
  {Osinde}, {Pagani}, {Pagano}, {Palaversa}, {Palicio}, {Panahi}, {Pawlak},
  {Pe{\~n}alosa Esteller}, {Penttil{\"a}}, {Piersimoni}, {Pineau}, {Plachy},
  {Plum}, {Poggio}, {Poretti}, {Poujoulet}, {Pr{\v{s}}a}, {Pulone}, {Racero},
  {Ragaini}, {Rainer}, {Raiteri}, {Rambaux}, {Ramos}, {Ramos-Lerate}, {Re
  Fiorentin}, {Regibo}, {Reyl{\'e}}, {Ripepi}, {Riva}, {Rixon}, {Robichon},
  {Robin}, {Roelens}, {Rohrbasser}, {Romero-G{\'o}mez}, {Rowell}, {Royer},
  {Rybicki}, {Sadowski}, {Sagrist{\`a} Sell{\'e}s}, {Sahlmann}, {Salgado},
  {Salguero}, {Samaras}, {Sanchez Gimenez}, {Sanna}, {Santove{\~n}a},
  {Sarasso}, {Schultheis}, {Sciacca}, {Segol}, {Segovia}, {S{\'e}gransan},
  {Semeux}, {Shahaf}, {Siddiqui}, {Siebert}, {Siltala}, {Slezak}, {Smart},
  {Solano}, {Solitro}, {Souami}, {Souchay}, {Spagna}, {Spoto}, {Steele},
  {Steidelm{\"u}ller}, {Stephenson}, {S{\"u}veges}, {Szabados}, {Szegedi-Elek},
  {Taris}, {Tauran}, {Taylor}, {Teixeira}, {Thuillot}, {Tonello}, {Torra},
  {Torra}, {Turon}, {Unger}, {Vaillant}, {van Dillen}, {Vanel}, {Vecchiato},
  {Viala}, {Vicente}, {Voutsinas}, {Weiler}, {Wevers}, {Wyrzykowski}, {Yoldas},
  {Yvard}, {Zhao}, {Zorec}, {Zucker}, {Zurbach}, \&
  {Zwitter}}]{2021A&A...649A...1G}
{Gaia Collaboration}, {Brown}, A.~G.~A., {Vallenari}, A., {et~al.} 2021, \aap,
  649, A1

\bibitem[{{Gaia Collaboration} {et~al.}(2016{\natexlab{a}}){Gaia
  Collaboration}, {Brown}, {Vallenari}, {Prusti}, {de Bruijne}, {Mignard},
  {Drimmel}, {Babusiaux}, {Bailer-Jones}, {Bastian}, {Biermann}, {Evans},
  {Eyer}, {Jansen}, {Jordi}, {Katz}, {Klioner}, {Lammers}, {Lindegren}, {Luri},
  {O'Mullane}, {Panem}, {Pourbaix}, {Randich}, {Sartoretti}, {Siddiqui},
  {Soubiran}, {Valette}, {van Leeuwen}, {Walton}, {Aerts}, {Arenou}, {Cropper},
  {H{\o}g}, {Lattanzi}, {Grebel}, {Holland}, {Huc}, {Passot}, {Perryman},
  {Bramante}, {Cacciari}, {Casta{\~n}eda}, {Chaoul}, {Cheek}, {De Angeli},
  {Fabricius}, {Guerra}, {Hern{\'a}ndez}, {Jean-Antoine-Piccolo}, {Masana},
  {Messineo}, {Mowlavi}, {Nienartowicz}, {Ord{\'o}{\~n}ez-Blanco}, {Panuzzo},
  {Portell}, {Richards}, {Riello}, {Seabroke}, {Tanga}, {Th{\'e}venin},
  {Torra}, {Els}, {Gracia-Abril}, {Comoretto}, {Garcia-Reinaldos}, {Lock},
  {Mercier}, {Altmann}, {Andrae}, {Astraatmadja}, {Bellas-Velidis}, {Benson},
  {Berthier}, {Blomme}, {Busso}, {Carry}, {Cellino}, {Clementini}, {Cowell},
  {Creevey}, {Cuypers}, {Davidson}, {De Ridder}, {de Torres}, {Delchambre},
  {Dell'Oro}, {Ducourant}, {Fr{\'e}mat}, {Garc{\'\i}a-Torres}, {Gosset},
  {Halbwachs}, {Hambly}, {Harrison}, {Hauser}, {Hestroffer}, {Hodgkin},
  {Huckle}, {Hutton}, {Jasniewicz}, {Jordan}, {Kontizas}, {Korn}, {Lanzafame},
  {Manteiga}, {Moitinho}, {Muinonen}, {Osinde}, {Pancino}, {Pauwels}, {Petit},
  {Recio-Blanco}, {Robin}, {Sarro}, {Siopis}, {Smith}, {Smith}, {Sozzetti},
  {Thuillot}, {van Reeven}, {Viala}, {Abbas}, {Abreu Aramburu}, {Accart},
  {Aguado}, {Allan}, {Allasia}, {Altavilla}, {{\'A}lvarez}, {Alves},
  {Anderson}, {Andrei}, {Anglada Varela}, {Antiche}, {Antoja}, {Ant{\'o}n},
  {Arcay}, {Bach}, {Baker}, {Balaguer-N{\'u}{\~n}ez}, {Barache}, {Barata},
  {Barbier}, {Barblan}, {Barrado y Navascu{\'e}s}, {Barros}, {Barstow},
  {Becciani}, {Bellazzini}, {Bello Garc{\'\i}a}, {Belokurov}, {Bendjoya},
  {Berihuete}, {Bianchi}, {Bienaym{\'e}}, {Billebaud}, {Blagorodnova},
  {Blanco-Cuaresma}, {Boch}, {Bombrun}, {Borrachero}, {Bouquillon}, {Bourda},
  {Bouy}, {Bragaglia}, {Breddels}, {Brouillet}, {Br{\"u}semeister},
  {Bucciarelli}, {Burgess}, {Burgon}, {Burlacu}, {Busonero}, {Buzzi}, {Caffau},
  {Cambras}, {Campbell}, {Cancelliere}, {Cantat-Gaudin}, {Carlucci},
  {Carrasco}, {Castellani}, {Charlot}, {Charnas}, {Chiavassa}, {Clotet},
  {Cocozza}, {Collins}, {Costigan}, {Crifo}, {Cross}, {Crosta}, {Crowley},
  {Dafonte}, {Damerdji}, {Dapergolas}, {David}, {David}, {De Cat}, {de Felice},
  {de Laverny}, {De Luise}, {De March}, {de Martino}, {de Souza}, {Debosscher},
  {del Pozo}, {Delbo}, {Delgado}, {Delgado}, {Di Matteo}, {Diakite},
  {Distefano}, {Dolding}, {Dos Anjos}, {Drazinos}, {Duran}, {Dzigan},
  {Edvardsson}, {Enke}, {Evans}, {Eynard Bontemps}, {Fabre}, {Fabrizio},
  {Faigler}, {Falc{\~a}o}, {Farr{\`a}s Casas}, {Federici}, {Fedorets},
  {Fern{\'a}ndez-Hern{\'a}ndez}, {Fernique}, {Fienga}, {Figueras}, {Filippi},
  {Findeisen}, {Fonti}, {Fouesneau}, {Fraile}, {Fraser}, {Fuchs}, {Gai},
  {Galleti}, {Galluccio}, {Garabato}, {Garc{\'\i}a-Sedano}, {Garofalo},
  {Garralda}, {Gavras}, {Gerssen}, {Geyer}, {Gilmore}, {Girona}, {Giuffrida},
  {Gomes}, {Gonz{\'a}lez-Marcos}, {Gonz{\'a}lez-N{\'u}{\~n}ez},
  {Gonz{\'a}lez-Vidal}, {Granvik}, {Guerrier}, {Guillout}, {Guiraud},
  {G{\'u}rpide}, {Guti{\'e}rrez-S{\'a}nchez}, {Guy}, {Haigron},
  {Hatzidimitriou}, {Haywood}, {Heiter}, {Helmi}, {Hobbs}, {Hofmann}, {Holl},
  {Holland}, {Hunt}, {Hypki}, {Icardi}, {Irwin}, {Jevardat de Fombelle},
  {Jofr{\'e}}, {Jonker}, {Jorissen}, {Julbe}, {Karampelas}, {Kochoska},
  {Kohley}, {Kolenberg}, {Kontizas}, {Koposov}, {Kordopatis}, {Koubsky},
  {Krone-Martins}, {Kudryashova}, {Kull}, {Bachchan}, {Lacoste-Seris}, {Lanza},
  {Lavigne}, {Le Poncin-Lafitte}, {Lebreton}, {Lebzelter}, {Leccia}, {Leclerc},
  {Lecoeur-Taibi}, {Lemaitre}, {Lenhardt}, {Leroux}, {Liao}, {Licata},
  {Lindstr{\o}m}, {Lister}, {Livanou}, {Lobel}, {L{\"o}ffler}, {L{\'o}pez},
  {Lorenz}, {MacDonald}, {Magalh{\~a}es Fernandes}, {Managau}, {Mann},
  {Mantelet}, {Marchal}, {Marchant}, {Marconi}, {Marinoni}, {Marrese},
  {Marschalk{\'o}}, {Marshall}, {Mart{\'\i}n-Fleitas}, {Martino}, {Mary},
  {Matijevi{\v{c}}}, {Mazeh}, {McMillan}, {Messina}, {Michalik}, {Millar},
  {Miranda}, {Molina}, {Molinaro}, {Molinaro}, {Moln{\'a}r}, {Moniez},
  {Montegriffo}, {Mor}, {Mora}, {Morbidelli}, {Morel}, {Morgenthaler},
  {Morris}, {Mulone}, {Muraveva}, {Musella}, {Narbonne}, {Nelemans},
  {Nicastro}, {Noval}, {Ord{\'e}novic}, {Ordieres-Mer{\'e}}, {Osborne},
  {Pagani}, {Pagano}, {Pailler}, {Palacin}, {Palaversa}, {Parsons}, {Pecoraro},
  {Pedrosa}, {Pentik{\"a}inen}, {Pichon}, {Piersimoni}, {Pineau}, {Plachy},
  {Plum}, {Poujoulet}, {Pr{\v{s}}a}, {Pulone}, {Ragaini}, {Rago}, {Rambaux},
  {Ramos-Lerate}, {Ranalli}, {Rauw}, {Read}, {Regibo}, {Reyl{\'e}}, {Ribeiro},
  {Rimoldini}, {Ripepi}, {Riva}, {Rixon}, {Roelens}, {Romero-G{\'o}mez},
  {Rowell}, {Royer}, {Ruiz-Dern}, {Sadowski}, {Sagrist{\`a} Sell{\'e}s},
  {Sahlmann}, {Salgado}, {Salguero}, {Sarasso}, {Savietto}, {Schultheis},
  {Sciacca}, {Segol}, {Segovia}, {Segransan}, {Shih}, {Smareglia}, {Smart},
  {Solano}, {Solitro}, {Sordo}, {Soria Nieto}, {Souchay}, {Spagna}, {Spoto},
  {Stampa}, {Steele}, {Steidelm{\"u}ller}, {Stephenson}, {Stoev}, {Suess},
  {S{\"u}veges}, {Surdej}, {Szabados}, {Szegedi-Elek}, {Tapiador}, {Taris},
  {Tauran}, {Taylor}, {Teixeira}, {Terrett}, {Tingley}, {Trager}, {Turon},
  {Ulla}, {Utrilla}, {Valentini}, {van Elteren}, {Van Hemelryck}, {van
  Leeuwen}, {Varadi}, {Vecchiato}, {Veljanoski}, {Via}, {Vicente}, {Vogt},
  {Voss}, {Votruba}, {Voutsinas}, {Walmsley}, {Weiler}, {Weingrill}, {Wevers},
  {Wyrzykowski}, {Yoldas}, {{\v{Z}}erjal}, {Zucker}, {Zurbach}, {Zwitter},
  {Alecu}, {Allen}, {Allende Prieto}, {Amorim}, {Anglada-Escud{\'e}},
  {Arsenijevic}, {Azaz}, {Balm}, {Beck}, {Bernstein}, {Bigot}, {Bijaoui},
  {Blasco}, {Bonfigli}, {Bono}, {Boudreault}, {Bressan}, {Brown}, {Brunet},
  {Bunclark}, {Buonanno}, {Butkevich}, {Carret}, {Carrion}, {Chemin},
  {Ch{\'e}reau}, {Corcione}, {Darmigny}, {de Boer}, {de Teodoro}, {de Zeeuw},
  {Delle Luche}, {Domingues}, {Dubath}, {Fodor}, {Fr{\'e}zouls}, {Fries},
  {Fustes}, {Fyfe}, {Gallardo}, {Gallegos}, {Gardiol}, {Gebran}, {Gomboc},
  {G{\'o}mez}, {Grux}, {Gueguen}, {Heyrovsky}, {Hoar}, {Iannicola}, {Isasi
  Parache}, {Janotto}, {Joliet}, {Jonckheere}, {Keil}, {Kim}, {Klagyivik},
  {Klar}, {Knude}, {Kochukhov}, {Kolka}, {Kos}, {Kutka}, {Lainey}, {LeBouquin},
  {Liu}, {Loreggia}, {Makarov}, {Marseille}, {Martayan}, {Martinez-Rubi},
  {Massart}, {Meynadier}, {Mignot}, {Munari}, {Nguyen}, {Nordlander}, {Ocvirk},
  {O'Flaherty}, {Olias Sanz}, {Ortiz}, {Osorio}, {Oszkiewicz}, {Ouzounis},
  {Palmer}, {Park}, {Pasquato}, {Peltzer}, {Peralta}, {P{\'e}turaud},
  {Pieniluoma}, {Pigozzi}, {Poels}, {Prat}, {Prod'homme}, {Raison}, {Rebordao},
  {Risquez}, {Rocca-Volmerange}, {Rosen}, {Ruiz-Fuertes}, {Russo}, {Sembay},
  {Serraller Vizcaino}, {Short}, {Siebert}, {Silva}, {Sinachopoulos}, {Slezak},
  {Soffel}, {Sosnowska}, {Strai{\v{z}}ys}, {ter Linden}, {Terrell}, {Theil},
  {Tiede}, {Troisi}, {Tsalmantza}, {Tur}, {Vaccari}, {Vachier}, {Valles}, {Van
  Hamme}, {Veltz}, {Virtanen}, {Wallut}, {Wichmann}, {Wilkinson}, {Ziaeepour},
  \& {Zschocke}}]{2016A&A...595A...2G}
{Gaia Collaboration}, {Brown}, A.~G.~A., {Vallenari}, A., {et~al.}
  2016{\natexlab{a}}, \aap, 595, A2

\bibitem[{{Gaia Collaboration} {et~al.}(2016{\natexlab{b}}){Gaia
  Collaboration}, {Prusti}, {de Bruijne}, {Brown}, {Vallenari}, {Babusiaux},
  {Bailer-Jones}, {Bastian}, {Biermann}, {Evans}, {Eyer}, {Jansen}, {Jordi},
  {Klioner}, {Lammers}, {Lindegren}, {Luri}, {Mignard}, {Milligan}, {Panem},
  {Poinsignon}, {Pourbaix}, {Randich}, {Sarri}, {Sartoretti}, {Siddiqui},
  {Soubiran}, {Valette}, {van Leeuwen}, {Walton}, {Aerts}, {Arenou}, {Cropper},
  {Drimmel}, {H{\o}g}, {Katz}, {Lattanzi}, {O'Mullane}, {Grebel}, {Holland},
  {Huc}, {Passot}, {Bramante}, {Cacciari}, {Casta{\~n}eda}, {Chaoul}, {Cheek},
  {De Angeli}, {Fabricius}, {Guerra}, {Hern{\'a}ndez}, {Jean-Antoine-Piccolo},
  {Masana}, {Messineo}, {Mowlavi}, {Nienartowicz}, {Ord{\'o}{\~n}ez-Blanco},
  {Panuzzo}, {Portell}, {Richards}, {Riello}, {Seabroke}, {Tanga},
  {Th{\'e}venin}, {Torra}, {Els}, {Gracia-Abril}, {Comoretto},
  {Garcia-Reinaldos}, {Lock}, {Mercier}, {Altmann}, {Andrae}, {Astraatmadja},
  {Bellas-Velidis}, {Benson}, {Berthier}, {Blomme}, {Busso}, {Carry},
  {Cellino}, {Clementini}, {Cowell}, {Creevey}, {Cuypers}, {Davidson}, {De
  Ridder}, {de Torres}, {Delchambre}, {Dell'Oro}, {Ducourant}, {Fr{\'e}mat},
  {Garc{\'\i}a-Torres}, {Gosset}, {Halbwachs}, {Hambly}, {Harrison}, {Hauser},
  {Hestroffer}, {Hodgkin}, {Huckle}, {Hutton}, {Jasniewicz}, {Jordan},
  {Kontizas}, {Korn}, {Lanzafame}, {Manteiga}, {Moitinho}, {Muinonen},
  {Osinde}, {Pancino}, {Pauwels}, {Petit}, {Recio-Blanco}, {Robin}, {Sarro},
  {Siopis}, {Smith}, {Smith}, {Sozzetti}, {Thuillot}, {van Reeven}, {Viala},
  {Abbas}, {Abreu Aramburu}, {Accart}, {Aguado}, {Allan}, {Allasia},
  {Altavilla}, {{\'A}lvarez}, {Alves}, {Anderson}, {Andrei}, {Anglada Varela},
  {Antiche}, {Antoja}, {Ant{\'o}n}, {Arcay}, {Atzei}, {Ayache}, {Bach},
  {Baker}, {Balaguer-N{\'u}{\~n}ez}, {Barache}, {Barata}, {Barbier}, {Barblan},
  {Baroni}, {Barrado y Navascu{\'e}s}, {Barros}, {Barstow}, {Becciani},
  {Bellazzini}, {Bellei}, {Bello Garc{\'\i}a}, {Belokurov}, {Bendjoya},
  {Berihuete}, {Bianchi}, {Bienaym{\'e}}, {Billebaud}, {Blagorodnova},
  {Blanco-Cuaresma}, {Boch}, {Bombrun}, {Borrachero}, {Bouquillon}, {Bourda},
  {Bouy}, {Bragaglia}, {Breddels}, {Brouillet}, {Br{\"u}semeister},
  {Bucciarelli}, {Budnik}, {Burgess}, {Burgon}, {Burlacu}, {Busonero}, {Buzzi},
  {Caffau}, {Cambras}, {Campbell}, {Cancelliere}, {Cantat-Gaudin}, {Carlucci},
  {Carrasco}, {Castellani}, {Charlot}, {Charnas}, {Charvet}, {Chassat},
  {Chiavassa}, {Clotet}, {Cocozza}, {Collins}, {Collins}, {Costigan}, {Crifo},
  {Cross}, {Crosta}, {Crowley}, {Dafonte}, {Damerdji}, {Dapergolas}, {David},
  {David}, {De Cat}, {de Felice}, {de Laverny}, {De Luise}, {De March}, {de
  Martino}, {de Souza}, {Debosscher}, {del Pozo}, {Delbo}, {Delgado},
  {Delgado}, {di Marco}, {Di Matteo}, {Diakite}, {Distefano}, {Dolding}, {Dos
  Anjos}, {Drazinos}, {Dur{\'a}n}, {Dzigan}, {Ecale}, {Edvardsson}, {Enke},
  {Erdmann}, {Escolar}, {Espina}, {Evans}, {Eynard Bontemps}, {Fabre},
  {Fabrizio}, {Faigler}, {Falc{\~a}o}, {Farr{\`a}s Casas}, {Faye}, {Federici},
  {Fedorets}, {Fern{\'a}ndez-Hern{\'a}ndez}, {Fernique}, {Fienga}, {Figueras},
  {Filippi}, {Findeisen}, {Fonti}, {Fouesneau}, {Fraile}, {Fraser}, {Fuchs},
  {Furnell}, {Gai}, {Galleti}, {Galluccio}, {Garabato}, {Garc{\'\i}a-Sedano},
  {Gar{\'e}}, {Garofalo}, {Garralda}, {Gavras}, {Gerssen}, {Geyer}, {Gilmore},
  {Girona}, {Giuffrida}, {Gomes}, {Gonz{\'a}lez-Marcos},
  {Gonz{\'a}lez-N{\'u}{\~n}ez}, {Gonz{\'a}lez-Vidal}, {Granvik}, {Guerrier},
  {Guillout}, {Guiraud}, {G{\'u}rpide}, {Guti{\'e}rrez-S{\'a}nchez}, {Guy},
  {Haigron}, {Hatzidimitriou}, {Haywood}, {Heiter}, {Helmi}, {Hobbs},
  {Hofmann}, {Holl}, {Holland}, {Hunt}, {Hypki}, {Icardi}, {Irwin}, {Jevardat
  de Fombelle}, {Jofr{\'e}}, {Jonker}, {Jorissen}, {Julbe}, {Karampelas},
  {Kochoska}, {Kohley}, {Kolenberg}, {Kontizas}, {Koposov}, {Kordopatis},
  {Koubsky}, {Kowalczyk}, {Krone-Martins}, {Kudryashova}, {Kull}, {Bachchan},
  {Lacoste-Seris}, {Lanza}, {Lavigne}, {Le Poncin-Lafitte}, {Lebreton},
  {Lebzelter}, {Leccia}, {Leclerc}, {Lecoeur-Taibi}, {Lemaitre}, {Lenhardt},
  {Leroux}, {Liao}, {Licata}, {Lindstr{\o}m}, {Lister}, {Livanou}, {Lobel},
  {L{\"o}ffler}, {L{\'o}pez}, {Lopez-Lozano}, {Lorenz}, {Loureiro},
  {MacDonald}, {Magalh{\~a}es Fernandes}, {Managau}, {Mann}, {Mantelet},
  {Marchal}, {Marchant}, {Marconi}, {Marie}, {Marinoni}, {Marrese},
  {Marschalk{\'o}}, {Marshall}, {Mart{\'\i}n-Fleitas}, {Martino}, {Mary},
  {Matijevi{\v{c}}}, {Mazeh}, {McMillan}, {Messina}, {Mestre}, {Michalik},
  {Millar}, {Miranda}, {Molina}, {Molinaro}, {Molinaro}, {Moln{\'a}r},
  {Moniez}, {Montegriffo}, {Monteiro}, {Mor}, {Mora}, {Morbidelli}, {Morel},
  {Morgenthaler}, {Morley}, {Morris}, {Mulone}, {Muraveva}, {Musella},
  {Narbonne}, {Nelemans}, {Nicastro}, {Noval}, {Ord{\'e}novic},
  {Ordieres-Mer{\'e}}, {Osborne}, {Pagani}, {Pagano}, {Pailler}, {Palacin},
  {Palaversa}, {Parsons}, {Paulsen}, {Pecoraro}, {Pedrosa}, {Pentik{\"a}inen},
  {Pereira}, {Pichon}, {Piersimoni}, {Pineau}, {Plachy}, {Plum}, {Poujoulet},
  {Pr{\v{s}}a}, {Pulone}, {Ragaini}, {Rago}, {Rambaux}, {Ramos-Lerate},
  {Ranalli}, {Rauw}, {Read}, {Regibo}, {Renk}, {Reyl{\'e}}, {Ribeiro},
  {Rimoldini}, {Ripepi}, {Riva}, {Rixon}, {Roelens}, {Romero-G{\'o}mez},
  {Rowell}, {Royer}, {Rudolph}, {Ruiz-Dern}, {Sadowski}, {Sagrist{\`a}
  Sell{\'e}s}, {Sahlmann}, {Salgado}, {Salguero}, {Sarasso}, {Savietto},
  {Schnorhk}, {Schultheis}, {Sciacca}, {Segol}, {Segovia}, {Segransan},
  {Serpell}, {Shih}, {Smareglia}, {Smart}, {Smith}, {Solano}, {Solitro},
  {Sordo}, {Soria Nieto}, {Souchay}, {Spagna}, {Spoto}, {Stampa}, {Steele},
  {Steidelm{\"u}ller}, {Stephenson}, {Stoev}, {Suess}, {S{\"u}veges}, {Surdej},
  {Szabados}, {Szegedi-Elek}, {Tapiador}, {Taris}, {Tauran}, {Taylor},
  {Teixeira}, {Terrett}, {Tingley}, {Trager}, {Turon}, {Ulla}, {Utrilla},
  {Valentini}, {van Elteren}, {Van Hemelryck}, {van Leeuwen}, {Varadi},
  {Vecchiato}, {Veljanoski}, {Via}, {Vicente}, {Vogt}, {Voss}, {Votruba},
  {Voutsinas}, {Walmsley}, {Weiler}, {Weingrill}, {Werner}, {Wevers},
  {Whitehead}, {Wyrzykowski}, {Yoldas}, {{\v{Z}}erjal}, {Zucker}, {Zurbach},
  {Zwitter}, {Alecu}, {Allen}, {Allende Prieto}, {Amorim},
  {Anglada-Escud{\'e}}, {Arsenijevic}, {Azaz}, {Balm}, {Beck}, {Bernstein},
  {Bigot}, {Bijaoui}, {Blasco}, {Bonfigli}, {Bono}, {Boudreault}, {Bressan},
  {Brown}, {Brunet}, {Bunclark}, {Buonanno}, {Butkevich}, {Carret}, {Carrion},
  {Chemin}, {Ch{\'e}reau}, {Corcione}, {Darmigny}, {de Boer}, {de Teodoro}, {de
  Zeeuw}, {Delle Luche}, {Domingues}, {Dubath}, {Fodor}, {Fr{\'e}zouls},
  {Fries}, {Fustes}, {Fyfe}, {Gallardo}, {Gallegos}, {Gardiol}, {Gebran},
  {Gomboc}, {G{\'o}mez}, {Grux}, {Gueguen}, {Heyrovsky}, {Hoar}, {Iannicola},
  {Isasi Parache}, {Janotto}, {Joliet}, {Jonckheere}, {Keil}, {Kim},
  {Klagyivik}, {Klar}, {Knude}, {Kochukhov}, {Kolka}, {Kos}, {Kutka}, {Lainey},
  {LeBouquin}, {Liu}, {Loreggia}, {Makarov}, {Marseille}, {Martayan},
  {Martinez-Rubi}, {Massart}, {Meynadier}, {Mignot}, {Munari}, {Nguyen},
  {Nordlander}, {Ocvirk}, {O'Flaherty}, {Olias Sanz}, {Ortiz}, {Osorio},
  {Oszkiewicz}, {Ouzounis}, {Palmer}, {Park}, {Pasquato}, {Peltzer}, {Peralta},
  {P{\'e}turaud}, {Pieniluoma}, {Pigozzi}, {Poels}, {Prat}, {Prod'homme},
  {Raison}, {Rebordao}, {Risquez}, {Rocca-Volmerange}, {Rosen}, {Ruiz-Fuertes},
  {Russo}, {Sembay}, {Serraller Vizcaino}, {Short}, {Siebert}, {Silva},
  {Sinachopoulos}, {Slezak}, {Soffel}, {Sosnowska}, {Strai{\v{z}}ys}, {ter
  Linden}, {Terrell}, {Theil}, {Tiede}, {Troisi}, {Tsalmantza}, {Tur},
  {Vaccari}, {Vachier}, {Valles}, {Van Hamme}, {Veltz}, {Virtanen}, {Wallut},
  {Wichmann}, {Wilkinson}, {Ziaeepour}, \& {Zschocke}}]{2016A&A...595A...1G}
{Gaia Collaboration}, {Prusti}, T., {de Bruijne}, J.~H.~J., {et~al.}
  2016{\natexlab{b}}, \aap, 595, A1

\bibitem[{{Ginski} {et~al.}(2021){Ginski}, {Mugrauer}, {Adam}, {Vogt}, \& {van
  Holstein}}]{2021A&A...649A.156G}
{Ginski}, C., {Mugrauer}, M., {Adam}, C., {Vogt}, N., \& {van Holstein}, R.~G.
  2021, \aap, 649, A156

\bibitem[{{Gravity Collaboration} {et~al.}(2017){Gravity Collaboration},
  {Abuter}, {Accardo}, {Amorim}, {Anugu}, {{\'A}vila}, {Azouaoui}, {Benisty},
  {Berger}, {Blind}, {Bonnet}, {Bourget}, {Brandner}, {Brast}, {Buron},
  {Burtscher}, {Cassaing}, {Chapron}, {Choquet}, {Cl{\'e}net}, {Collin},
  {Coud{\'e} Du Foresto}, {de Wit}, {de Zeeuw}, {Deen},
  {Delplancke-Str{\"o}bele}, {Dembet}, {Derie}, {Dexter}, {Duvert}, {Ebert},
  {Eckart}, {Eisenhauer}, {Esselborn}, {F{\'e}dou}, {Finger}, {Garcia}, {Garcia
  Dabo}, {Garcia Lopez}, {Gendron}, {Genzel}, {Gillessen}, {Gonte}, {Gordo},
  {Grould}, {Gr{\"o}zinger}, {Guieu}, {Haguenauer}, {Hans}, {Haubois}, {Haug},
  {Haussmann}, {Henning}, {Hippler}, {Horrobin}, {Huber}, {Hubert}, {Hubin},
  {Hummel}, {Jakob}, {Janssen}, {Jochum}, {Jocou}, {Kaufer}, {Kellner},
  {Kendrew}, {Kern}, {Kervella}, {Kiekebusch}, {Klein}, {Kok}, {Kolb}, {Kulas},
  {Lacour}, {Lapeyr{\`e}re}, {Lazareff}, {Le Bouquin}, {L{\`e}na}, {Lenzen},
  {L{\'e}v{\^e}que}, {Lippa}, {Magnard}, {Mehrgan}, {Mellein}, {M{\'e}rand},
  {Moreno-Ventas}, {Moulin}, {M{\"u}ller}, {M{\"u}ller}, {Neumann}, {Oberti},
  {Ott}, {Pallanca}, {Panduro}, {Pasquini}, {Paumard}, {Percheron}, {Perraut},
  {Perrin}, {Pfl{\"u}ger}, {Pfuhl}, {Phan Duc}, {Plewa}, {Popovic}, {Rabien},
  {Ram{\'\i}rez}, {Ramos}, {Rau}, {Riquelme}, {Rohloff}, {Rousset},
  {Sanchez-Bermudez}, {Scheithauer}, {Sch{\"o}ller}, {Schuhler}, {Spyromilio},
  {Straubmeier}, {Sturm}, {Suarez}, {Tristram}, {Ventura}, {Vincent},
  {Waisberg}, {Wank}, {Weber}, {Wieprecht}, {Wiest}, {Wiezorrek}, {Wittkowski},
  {Woillez}, {Wolff}, {Yazici}, {Ziegler}, \& {Zins}}]{2017A&A...602A..94G}
{Gravity Collaboration}, {Abuter}, R., {Accardo}, M., {et~al.} 2017, \aap, 602,
  A94

\bibitem[{{Gravity Collaboration} {et~al.}(2019){Gravity Collaboration},
  {Lacour}, {Nowak}, {Wang}, {Pfuhl}, {Eisenhauer}, {Abuter}, {Amorim},
  {Anugu}, {Benisty}, {Berger}, {Beust}, {Blind}, {Bonnefoy}, {Bonnet},
  {Bourget}, {Brandner}, {Buron}, {Collin}, {Charnay}, {Chapron}, {Cl{\'e}net},
  {Coud{\'e} Du Foresto}, {de Zeeuw}, {Deen}, {Dembet}, {Dexter}, {Duvert},
  {Eckart}, {F{\"o}rster Schreiber}, {F{\'e}dou}, {Garcia}, {Garcia Lopez},
  {Gao}, {Gendron}, {Genzel}, {Gillessen}, {Gordo}, {Greenbaum}, {Habibi},
  {Haubois}, {Hau{\ss}mann}, {Henning}, {Hippler}, {Horrobin}, {Hubert},
  {Jimenez Rosales}, {Jocou}, {Kendrew}, {Kervella}, {Kolb}, {Lagrange},
  {Lapeyr{\`e}re}, {Le Bouquin}, {L{\'e}na}, {Lippa}, {Lenzen}, {Maire},
  {Molli{\`e}re}, {Ott}, {Paumard}, {Perraut}, {Perrin}, {Pueyo}, {Rabien},
  {Ram{\'\i}rez}, {Rau}, {Rodr{\'\i}guez-Coira}, {Rousset}, {Sanchez-Bermudez},
  {Scheithauer}, {Schuhler}, {Straub}, {Straubmeier}, {Sturm}, {Tacconi},
  {Vincent}, {van Dishoeck}, {von Fellenberg}, {Wank}, {Waisberg}, {Widmann},
  {Wieprecht}, {Wiest}, {Wiezorrek}, {Woillez}, {Yazici}, {Ziegler}, \&
  {Zins}}]{2019A&A...623L..11G}
{Gravity Collaboration}, {Lacour}, S., {Nowak}, M., {et~al.} 2019, \aap, 623,
  L11

\bibitem[{Gregory(2005)}]{gregory2005bayesian}
Gregory, P. 2005, The Astrophysical Journal, 631, 1198

\bibitem[{Gregory(2011)}]{gregory2011bayesian}
Gregory, P.~C. 2011, Monthly Notices of the Royal Astronomical Society, 410, 94

\bibitem[{{Hatzes} {et~al.}(2003){Hatzes}, {Cochran}, {Endl}, {McArthur},
  {Paulson}, {Walker}, {Campbell}, \& {Yang}}]{2003ApJ...599.1383H}
{Hatzes}, A.~P., {Cochran}, W.~D., {Endl}, M., {et~al.} 2003, \apj, 599, 1383

\bibitem[{{Heidari} {et~al.}(2022){Heidari}, {Boisse}, {Orell-Miquel},
  {H{\'e}brard}, {Acu{\~n}a}, {Hara}, {Lillo-Box}, {Eastman}, {Arnold},
  {Astudillo-Defru}, {Adibekyan}, {Bieryla}, {Bonfils}, {Bouchy}, {Barclay},
  {Brasseur}, {Borgniet}, {Bourrier}, {Buchhave}, {Behmard}, {Beard},
  {Batalha}, {Courcol}, {Cort{\'e}s-Zuleta}, {Collins}, {Carmona},
  {Crossfield}, {Chontos}, {Delfosse}, {Dalal}, {Deleuil}, {Demangeon},
  {D{\'\i}az}, {Dumusque}, {Daylan}, {Dragomir}, {Delgado Mena}, {Dressing},
  {Dai}, {Dalba}, {Ehrenreich}, {Forveille}, {Fulton}, {Fetherolf},
  {Gaisn{\'e}}, {Giacalone}, {Riazi}, {Hoyer}, {Hobson}, {Howard}, {Huber},
  {Hill}, {Hirsch}, {Isaacson}, {Jenkins}, {Kane}, {Kiefer}, {Luque}, {Latham},
  {Lubin}, {Lopez}, {Mousis}, {Moutou}, {Montagnier}, {Mignon}, {Mayo},
  {Mo{\v{c}}nik}, {Murphy}, {Palle}, {Pepe}, {Petigura}, {Rey}, {Ricker},
  {Robertson}, {Roy}, {Rubenzahl}, {Rosenthal}, {Santerne}, {Santos}, {Sousa},
  {Stassun}, {Stalport}, {Scarsdale}, {Str{\o}m}, {Seager}, {Segransan},
  {Tenenbaum}, {Tronsgaard}, {Udry}, {Vanderspek}, {Vakili}, {Winn}, \&
  {Weiss}}]{2022A&A...658A.176H}
{Heidari}, N., {Boisse}, I., {Orell-Miquel}, J., {et~al.} 2022, \aap, 658, A176

\bibitem[{Hoffman {et~al.}(2014)Hoffman, Gelman, {et~al.}}]{hoffman2014no}
Hoffman, M.~D., Gelman, A., {et~al.} 2014, J. Mach. Learn. Res., 15, 1593

\bibitem[{Hou {et~al.}(2012)Hou, Goodman, Hogg, Weare, \&
  Schwab}]{hou2012affine}
Hou, F., Goodman, J., Hogg, D.~W., Weare, J., \& Schwab, C. 2012, The
  Astrophysical Journal, 745, 198

\bibitem[{{Jang-Condell}(2015)}]{2015ApJ...799..147J}
{Jang-Condell}, H. 2015, \apj, 799, 147

\bibitem[{{Jenkins} {et~al.}(2011){Jenkins}, {Murgas}, {Rojo}, {Jones},
  {Day-Jones}, {Jones}, {Clarke}, {Ruiz}, \& {Pinfield}}]{2011A&A...531A...8J}
{Jenkins}, J.~S., {Murgas}, F., {Rojo}, P., {et~al.} 2011, \aap, 531, A8

\bibitem[{{Kervella} {et~al.}(2019){Kervella}, {Arenou}, {Mignard}, \&
  {Th{\'e}venin}}]{2019A&A...623A..72K}
{Kervella}, P., {Arenou}, F., {Mignard}, F., \& {Th{\'e}venin}, F. 2019, \aap,
  623, A72

\bibitem[{{Kervella} {et~al.}(2022){Kervella}, {Arenou}, \&
  {Th{\'e}venin}}]{2022A&A...657A...7K}
{Kervella}, P., {Arenou}, F., \& {Th{\'e}venin}, F. 2022, \aap, 657, A7

\bibitem[{{Konacki}(2005)}]{2005AAS...207.8402K}
{Konacki}, M. 2005, in American Astronomical Society Meeting Abstracts, Vol.
  207, American Astronomical Society Meeting Abstracts, 84.02

\bibitem[{{Kozai}(1962)}]{1962AJ.....67..591K}
{Kozai}, Y. 1962, \aj, 67, 591

\bibitem[{{Krymolowski} \& {Mazeh}(1999)}]{1999MNRAS.304..720K}
{Krymolowski}, Y. \& {Mazeh}, T. 1999, \mnras, 304, 720

\bibitem[{{Langlois} {et~al.}(2021){Langlois}, {Gratton}, {Lagrange},
  {Delorme}, {Boccaletti}, {Bonnefoy}, {Maire}, {Mesa}, {Chauvin}, {Desidera},
  {Vigan}, {Cheetham}, {Hagelberg}, {Feldt}, {Meyer}, {Rubini}, {Le Coroller},
  {Cantalloube}, {Biller}, {Bonavita}, {Bhowmik}, {Brandner}, {Daemgen},
  {D'Orazi}, {Flasseur}, {Fontanive}, {Galicher}, {Girard}, {Janin-Potiron},
  {Janson}, {Keppler}, {Kopytova}, {Lagadec}, {Lannier}, {Lazzoni}, {Ligi},
  {Meunier}, {Perreti}, {Perrot}, {Rodet}, {Romero}, {Rouan}, {Samland},
  {Salter}, {Sissa}, {Schmidt}, {Zurlo}, {Mouillet}, {Denis}, {Thiebaut},
  {Milli}, {Wahhaj}, {Beuzit}, {Dominik}, {Henning}, {Menard}, {Muller},
  {Schmid}, {Turatto}, {Udry}, {Abe}, {Antichi}, {Allard}, {Baruffolo},
  {Baudoz}, {Baudrand}, {Bazzon}, {Blanchard}, {Carbillet}, {Carle}, {Cascone},
  {Charton}, {Claudi}, {Costille}, {De Caprio}, {Delboulbe}, {Dohlen},
  {Fantinel}, {Feautrier}, {Fusco}, {Gigan}, {Giro}, {Gisler}, {Gluck}, {Gry},
  {Hubin}, {Hugot}, {Jaquet}, {Kasper}, {Le Mignant}, {Llored}, {Madec},
  {Magnard}, {Martinez}, {Maurel}, {Messina}, {Moller-Nilsson}, {Mugnier},
  {Moulin}, {Origne}, {Pavlov}, {Perret}, {Petit}, {Pragt}, {Puget}, {Rabou},
  {Ramos}, {Rigal}, {Rochat}, {Roelfsema}, {Rousset}, {Roux}, {Salasnich},
  {Sauvage}, {Sevin}, {Soenke}, {Stadler}, {Suarez}, {Weber}, {Wildi}, \&
  {Rickman}}]{Langlois2021}
{Langlois}, M., {Gratton}, R., {Lagrange}, A.~M., {et~al.} 2021, arXiv
  e-prints, arXiv:2103.03976

\bibitem[{{Lapeyrere} {et~al.}(2014){Lapeyrere}, {Kervella}, {Lacour},
  {Azouaoui}, {Garcia-Dabo}, {Perrin}, {Eisenhauer}, {Perraut}, {Straubmeier},
  {Amorim}, \& {Brandner}}]{2014SPIE.9146E..2DL}
{Lapeyrere}, V., {Kervella}, P., {Lacour}, S., {et~al.} 2014, in Society of
  Photo-Optical Instrumentation Engineers (SPIE) Conference Series, Vol. 9146,
  Optical and Infrared Interferometry IV, ed. J.~K. {Rajagopal}, M.~J.
  {Creech-Eakman}, \& F.~{Malbet}, 91462D

\bibitem[{{Law} {et~al.}(2014){Law}, {Morton}, {Baranec}, {Riddle},
  {Ravichandran}, {Ziegler}, {Johnson}, {Tendulkar}, {Bui}, {Burse}, {Das},
  {Dekany}, {Kulkarni}, {Punnadi}, \& {Ramaprakash}}]{2014ApJ...791...35L}
{Law}, N.~M., {Morton}, T., {Baranec}, C., {et~al.} 2014, \apj, 791, 35

\bibitem[{Liu \& Nocedal(1989)}]{liu1989limited}
Liu, D.~C. \& Nocedal, J. 1989, Mathematical programming, 45, 503

\bibitem[{{Luhman} \& {Jayawardhana}(2002)}]{2002ApJ...566.1132L}
{Luhman}, K.~L. \& {Jayawardhana}, R. 2002, \apj, 566, 1132

\bibitem[{{Marzari} \& {Barbieri}(2007)}]{2007A&A...467..347M}
{Marzari}, F. \& {Barbieri}, M. 2007, \aap, 467, 347

\bibitem[{{Michel} \& {Mugrauer}(2021)}]{2021FrASS...8...14M}
{Michel}, K.~U. \& {Mugrauer}, M. 2021, Frontiers in Astronomy and Space
  Sciences, 8, 14

\bibitem[{{Mordasini} {et~al.}(2008){Mordasini}, {Alibert}, {Benz}, \&
  {Naef}}]{2008ASPC..398..235M}
{Mordasini}, C., {Alibert}, Y., {Benz}, W., \& {Naef}, D. 2008, in Astronomical
  Society of the Pacific Conference Series, Vol. 398, Extreme Solar Systems,
  ed. D.~{Fischer}, F.~A. {Rasio}, S.~E. {Thorsett}, \& A.~{Wolszczan}, 235

\bibitem[{{Mugrauer}(2019)}]{2019MNRAS.490.5088M}
{Mugrauer}, M. 2019, \mnras, 490, 5088

\bibitem[{{Mugrauer} \& {Ginski}(2015)}]{2015MNRAS.450.3127M}
{Mugrauer}, M. \& {Ginski}, C. 2015, \mnras, 450, 3127

\bibitem[{Nelson {et~al.}(2013)Nelson, Ford, \& Payne}]{nelson2013run}
Nelson, B., Ford, E.~B., \& Payne, M.~J. 2013, The Astrophysical Journal
  Supplement Series, 210, 11

\bibitem[{{Paardekooper} {et~al.}(2008){Paardekooper}, {Th{\'e}bault}, \&
  {Mellema}}]{2008MNRAS.386..973P}
{Paardekooper}, S.~J., {Th{\'e}bault}, P., \& {Mellema}, G. 2008, \mnras, 386,
  973

\bibitem[{{Patience} {et~al.}(2002){Patience}, {White}, {Ghez}, {McCabe},
  {McLean}, {Larkin}, {Prato}, {Kim}, {Lloyd}, {Liu}, {Graham}, {Macintosh},
  {Gavel}, {Max}, {Bauman}, {Olivier}, {Wizinowich}, \&
  {Acton}}]{2002ApJ...581..654P}
{Patience}, J., {White}, R.~J., {Ghez}, A.~M., {et~al.} 2002, \apj, 581, 654

\bibitem[{{Queloz} {et~al.}(2000){Queloz}, {Mayor}, {Weber}, {Bl{\'e}cha},
  {Burnet}, {Confino}, {Naef}, {Pepe}, {Santos}, \&
  {Udry}}]{2000A&A...354...99Q}
{Queloz}, D., {Mayor}, M., {Weber}, L., {et~al.} 2000, \aap, 354, 99

\bibitem[{{Raghavan} {et~al.}(2006){Raghavan}, {Henry}, {Mason}, {Subasavage},
  {Jao}, {Beaulieu}, \& {Hambly}}]{2006ApJ...646..523R}
{Raghavan}, D., {Henry}, T.~J., {Mason}, B.~D., {et~al.} 2006, \apj, 646, 523

\bibitem[{{Silsbee} \& {Rafikov}(2021)}]{2021A&A...652A.104S}
{Silsbee}, K. \& {Rafikov}, R.~R. 2021, \aap, 652, A104

\bibitem[{{Tamuz} {et~al.}(2008){Tamuz}, {S{\'e}gransan}, {Udry}, {Mayor},
  {Eggenberger}, {Naef}, {Pepe}, {Queloz}, {Santos}, {Demory}, {Figuera},
  {Marmier}, \& {Montagnier}}]{2008A&A...480L..33T}
{Tamuz}, O., {S{\'e}gransan}, D., {Udry}, S., {et~al.} 2008, \aap, 480, L33

\bibitem[{{Thebault}(2011)}]{2011CeMDA.111...29T}
{Thebault}, P. 2011, Celestial Mechanics and Dynamical Astronomy, 111, 29

\bibitem[{{Thebault} \& {Haghighipour}(2015)}]{2015pes..book..309T}
{Thebault}, P. \& {Haghighipour}, N. 2015, {Planet Formation in Binaries},
  309--340

\bibitem[{{Th{\'e}bault} {et~al.}(2006){Th{\'e}bault}, {Marzari}, \&
  {Scholl}}]{2006Icar..183..193T}
{Th{\'e}bault}, P., {Marzari}, F., \& {Scholl}, H. 2006, \icarus, 183, 193

\bibitem[{{Tokovinin}(2018)}]{Toko2018}
{Tokovinin}, A. 2018, \pasp, 130, 035002

\bibitem[{{Tokovinin} \& {Cantarutti}(2008)}]{tokovinin2008}
{Tokovinin}, A. \& {Cantarutti}, R. 2008, \pasp, 120, 170

\bibitem[{{Tokovinin} {et~al.}(2010){Tokovinin}, {Mason}, \&
  {Hartkopf}}]{TokMasHar2010}
{Tokovinin}, A., {Mason}, B.~D., \& {Hartkopf}, W.~I. 2010, \aj, 139, 743

\bibitem[{{Tokovinin} {et~al.}(2021){Tokovinin}, {Mason}, {Mendez}, {Costa},
  {Mann}, \& {Henry}}]{2021AJ....162...41T}
{Tokovinin}, A., {Mason}, B.~D., {Mendez}, R.~A., {et~al.} 2021, \aj, 162, 41

\bibitem[{{Toyota} {et~al.}(2009){Toyota}, {Itoh}, {Ishiguma}, {Urakawa},
  {Murata}, {Oasa}, {Matsuyama}, {Funayama}, {Sato}, \&
  {Mukai}}]{2009PASJ...61...19T}
{Toyota}, E., {Itoh}, Y., {Ishiguma}, S., {et~al.} 2009, \pasj, 61, 19

\bibitem[{Videla {et~al.}(2022)Videla, Mendez, Claver{\'\i}a, Silva, \&
  Orchard}]{videla2022bayesian}
Videla, M., Mendez, R.~A., Claver{\'\i}a, R.~M., Silva, J.~F., \& Orchard,
  M.~E. 2022, The Astronomical Journal, 163, 220

\bibitem[{{Vigan} {et~al.}(2010){Vigan}, {Moutou}, {Langlois}, {Allard},
  {Boccaletti}, {Carbillet}, {Mouillet}, \& {Smith}}]{Vigan2010dbi}
{Vigan}, A., {Moutou}, C., {Langlois}, M., {et~al.} 2010, \mnras, 407, 71

\bibitem[{{Villegas} {et~al.}(2021){Villegas}, {Mendez}, {Silva}, \&
  {Orchard}}]{2021PASP..133g4501V}
{Villegas}, C., {Mendez}, R.~A., {Silva}, J.~F., \& {Orchard}, M.~E. 2021,
  \pasp, 133, 074501

\bibitem[{{Wright} {et~al.}(2004){Wright}, {Marcy}, {Butler}, \&
  {Vogt}}]{2004ApJS..152..261W}
{Wright}, J.~T., {Marcy}, G.~W., {Butler}, R.~P., \& {Vogt}, S.~S. 2004, \apjs,
  152, 261

\bibitem[{{Ziegler} {et~al.}(2021){Ziegler}, {Tokovinin}, {Latiolais},
  {Brice{\~n}o}, {Law}, \& {Mann}}]{2021AJ....162..192Z}
{Ziegler}, C., {Tokovinin}, A., {Latiolais}, M., {et~al.} 2021, \aj, 162, 192

\bibitem[{{Zucker} \& {Mazeh}(2002)}]{2002ApJ...568L.113Z}
{Zucker}, S. \& {Mazeh}, T. 2002, \apjl, 568, L113

\end{thebibliography}
